\documentclass[useAMS,usenatbib]{mn2e}

\usepackage{graphicx}

\def\jh{\mbox{$(J-H)$}}

\def\jk{\mbox{$(J-K_s)$}}

\def\ebv{\mbox{$E(B-V)$}}

\def\ejh{\mbox{$E(J-H)$}}

\def\rc{\mbox{$R_{\rm c}$}}
\def\rl{\mbox{$R_{\rm l}$}}
\def\rx{\mbox{$R_{\rm ext}$}}
\def\rt{\mbox{$R_{\rm t}$}}

\def\ms{\mbox{$M_\odot$}}
\def\ds{\mbox{$d_\odot$}}
\def\rs{\mbox{$R_\odot$}}
\def\dgc{\mbox{$R_{\rm GC}$}}
\def\mobs{\mbox{$m_{obs}$}}
\def\jj{\mbox{$J$}}
\def\hh{\mbox{$H$}}
\def\ks{\mbox{$K_s$}}

\def\ns{\mbox{$\rm N_{1\sigma}$}}
\def\no{\mbox{$\rm N_{obs}$}}
\def\nc{\mbox{$\rm N_{cl}$}}
\def\sFS{\mbox{$\rm\sigma_{FS}$}}
\def\fsU{\mbox{$\rm FS_{unif}$}}
\def\mTO{\mbox{$\rm m_{TO}$}}

\title[A new open cluster with a possible PN]{Discovery of an open cluster with a possible
physical association with a planetary nebula}

\author[C. Bonatto, E. Bica and J.F.C. Santos Jr.]{C. Bonatto$^1$, E. Bica$^1$ and J.F.C. Santos Jr.$^2$\\
$^1$ Departamento de Astronomia, Universidade Federal do Rio Grande do Sul, Av. Bento Gon\c{c}alves 9500\\
Porto Alegre 91501-970, RS, Brazil\\
$^2$ Departamento de F\'\i sica, ICEx, Universidade Federal de Minas Gerais, Av. Ant\^onio Carlos 6627\\
Belo Horizonte 30123-970, MG, Brazil \\ }

\begin{document}

\pagerange{\pageref{firstpage}--\pageref{lastpage}} 

\maketitle

\label{firstpage}

\begin{abstract}
We report the discovery of a new open cluster (OC) in the Galaxy at $\ell=167.0^\circ$ 
and $b=-1.0^\circ$. Its field includes the planetary nebula (PN) PK\,167-0.1.
We study the possible associations of the PN/OC pairs 
NGC\,2818/NGC\,2818A, NGC\,2438/M\,46 (NGC\,2437), PK\,6+2.5/NGC\,6469, as well as of
the PN PK\,167-0.1 with New\,Cluster\,1. The analyses are based 
on near-infrared colour-magnitude diagrams (CMDs) and stellar radial density profiles (RDPs). 
NGC\,6469 is located in a heavily contaminated bulge field.  The CMD
morphology, especially for the latter two cases, is defined with a field star decontamination 
algorithm applied to the 2MASS \jj, \hh, and \ks\ photometry. Field decontamination for the
OCs NGC\,2818A and M\,46 produced better defined CMDs and more accurate cluster parameters 
than in the literature. Those pieces of evidence point to M\,46 as physically associated  
with the PN NGC\,2438. The same occurs for the OC NGC\,2818A and the PN NGC\,2818, however 
previous radial velocity arguments indicate that they are not associated.
The OC NGC\,6469 does not appear to be associated with the PN PK\,6+2.5, which probably belongs
to the bulge. Finally, the distance of the OC New\,Cluster\,1 is consistent with a physical 
association with the PN PK\,167-0.1. 
\end{abstract}

\begin{keywords}
{\em (Galaxy:)} open clusters and associations: general; {\em (ISM:)} planetary nebulae: 
general.

\end{keywords}

\section{Introduction}
\label{Intro}

Planetary nebulae (PNe) are late stages in stellar evolution, occurring for stars with mass in 
the range 1 --- 6.5\,\ms (\citealt{Weidemann00}), possibly with the upper limit at $\sim8\,\ms$.
Most of the difficulty in understanding such stages that lead to PNe is associated with the 
determination of the PN distance and its progenitor mass. {\bf Probably, the largest source
of error is the distance which, in most cases, is known with a $\sim50\%$ uncertainty
(e.g. \citealt{Zhang95}), or more. However, both problems can be minimised
if the PN is physically associated with a star cluster. Star cluster distances, in general,
can be determined with a high precision and, in the case of physical association, the PN progenitor
mass can be assumed to be $\approx10\%$ larger than that of the turnoff (TO), stars referred to by
\citet{Meynet93} as red-turnoff objects. Exceptions are the cases where mass transfer from a close
binary is involved. In fact, recent radial velocity measurements, although restricted to relatively
small samples, suggest than more than 90\% of the central stars in PNe might have companions (e.g.
\citealt{deM06}).}

Since the early observation of the PN NGC\,2438, which is surrounded by stars of the open
cluster (OC) M\,46 (NGC\,2437), photometric and spectroscopic methods have been developed to test
physical associations between such objects. \citet{Ziz75} published a list of 10 OCs with nearby
projected PNe. \citet{MTL07} provided a list of 13 possible physical associations and an additional
list of 17 angular coincidences ($\Delta\,R<15\arcmin$). They also present physical criteria for
membership based on similarities in velocity, reddening, and the ratio of estimated distances.

\citet{MTL07} reviewed theoretical and observational aspects of stellar evolution to impose limiting
parameters for PNe to be OC members. Many coincidences can be discarded if the PN is a bulge member
(\citealt{Ziz75}), as indicated by position along the disk and radial velocities. OCs younger than
28\,Myr produce type II SN. The short-lived PN phase varies from $10^3$ to $10^5$ years, according to
the progenitor mass (\citealt{KA2000}, \citealt{SB96}), and does not favour associations. PNe observed
in globular clusters (e.g. \citealt{KA2000}) imply that their turnoffs at $\la1\,\ms$ are below the
lower mass limit for PN progenitors. This invokes mass transfer in binary systems to produce PN in 
globular clusters. \citet{deM06}, \citet{Soker2006}, and \citet{Zijlstra07} presented a scenario where
most PNe arise from binaries. Consequently, care is necessary because the cluster age may not correspond 
to the PN mass progenitor.

Among the list of 30 PN/OC coincidences provided by \citet{MTL07}, we selected the PN/OC pairs
NGC\,2818/NGC\,2818A and NGC\,2438/M\,46 (NGC\,2437). We also include PK\,6+2.5/NGC\,6469 and
PK\,167-0.1, which is projected in the field of an as-yet unknown star cluster. We name this
low-contrast cluster New\,Cluster\,1. In the present paper we derive accurate distances to these
OCs, which in turn, can be used to investigate the possible associations.

\begin{figure*}
\begin{minipage}[b]{0.50\linewidth}
\includegraphics[width=\textwidth]{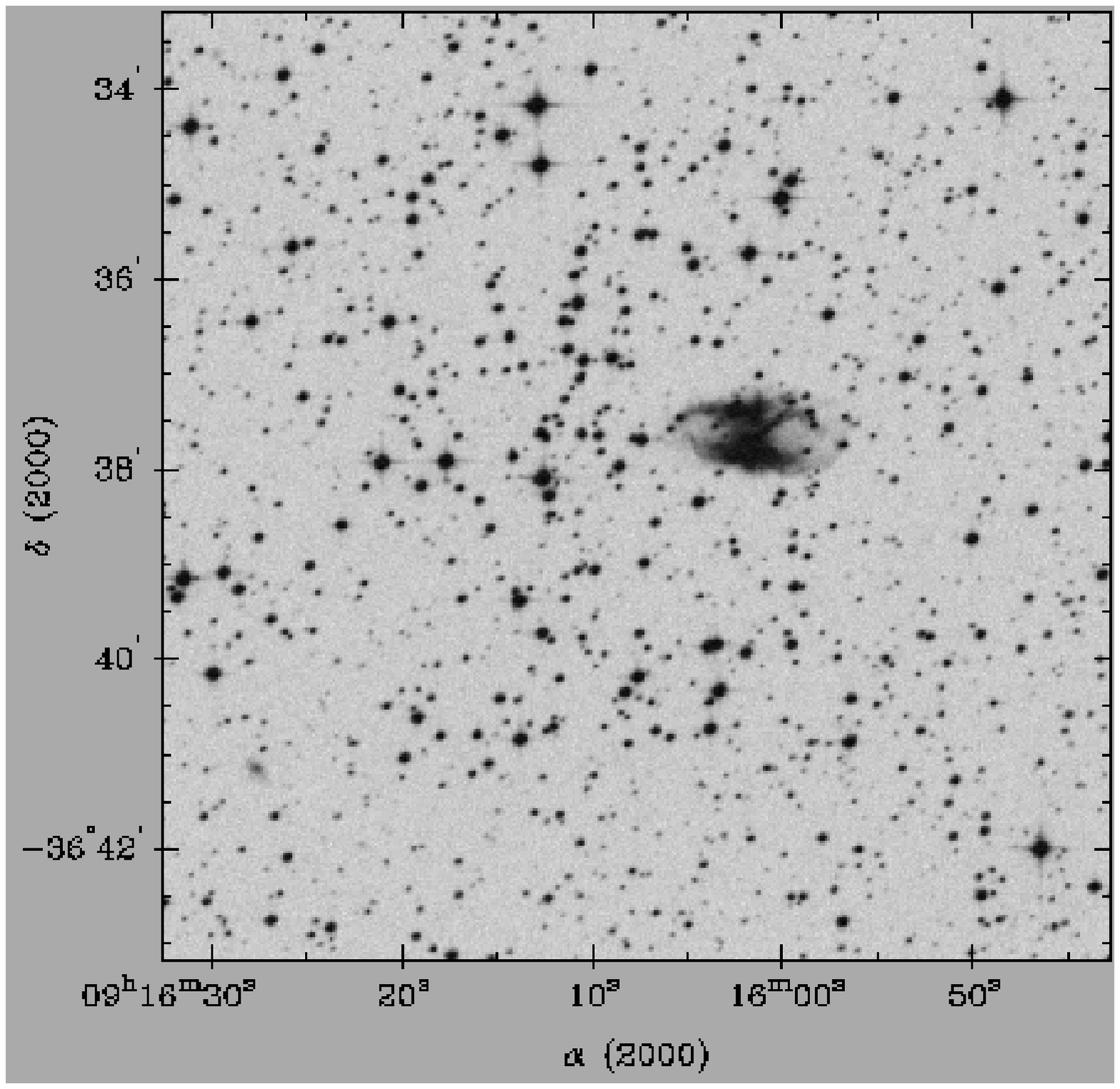}
\end{minipage}\hfill
\begin{minipage}[b]{0.50\linewidth}
\includegraphics[width=\textwidth]{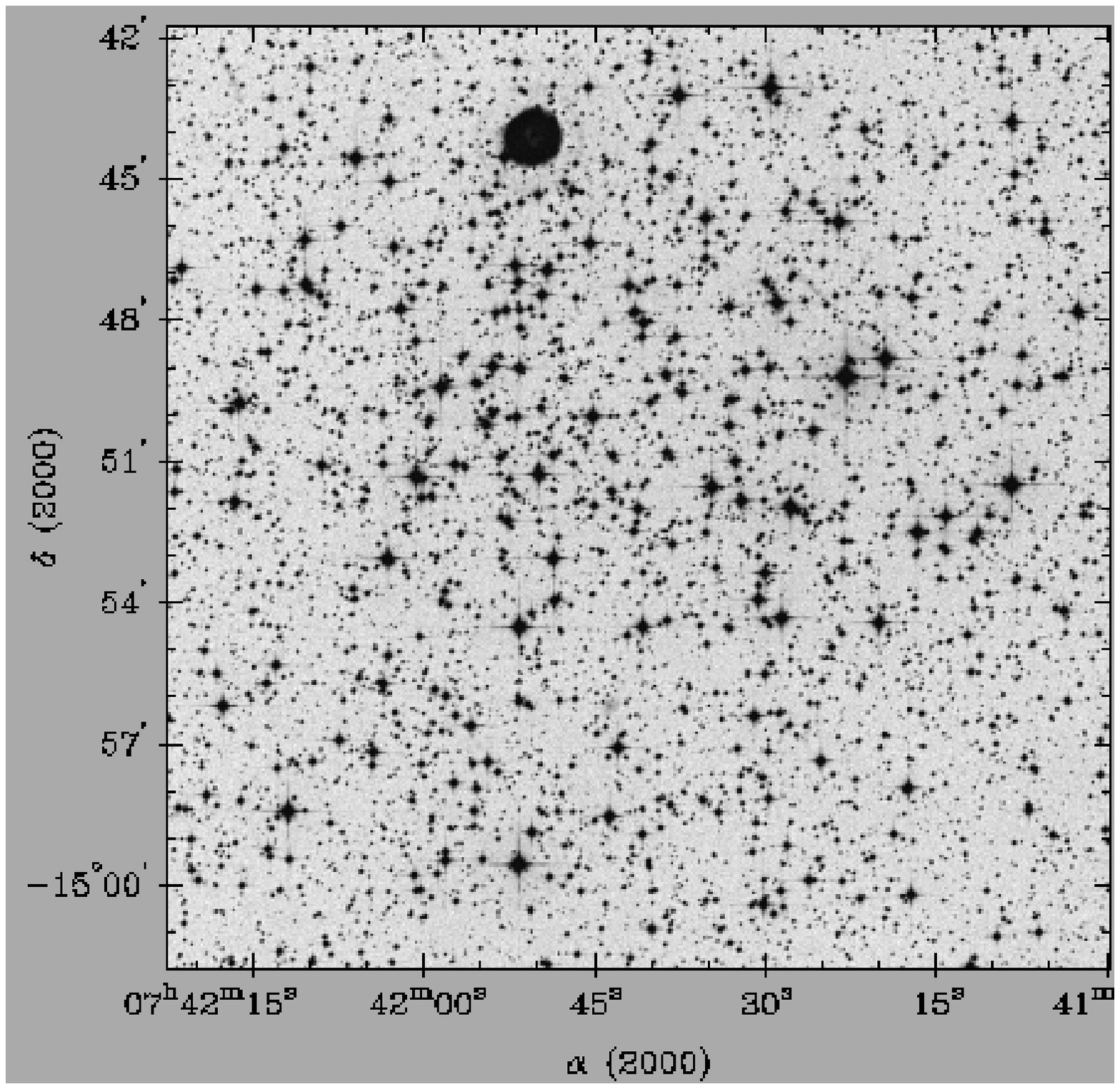}
\end{minipage}\hfill
\caption[]{Left panel: $10\arcmin\times10\arcmin$ XDSS R image of the OC NGC\,2818A and the PN NGC\,2818.
Right panel: $20\arcmin\times20\arcmin$ XDSS R image of the OC M\,46 and the PN NGC\,2438. Images centred
on the 2MASS coordinates (cols.~5 and 6 of Table~\ref{tab1}). North to the top and East to the left.}
\label{fig1}
\end{figure*}

\begin{figure*}
\begin{minipage}[b]{0.50\linewidth}
\includegraphics[width=\textwidth]{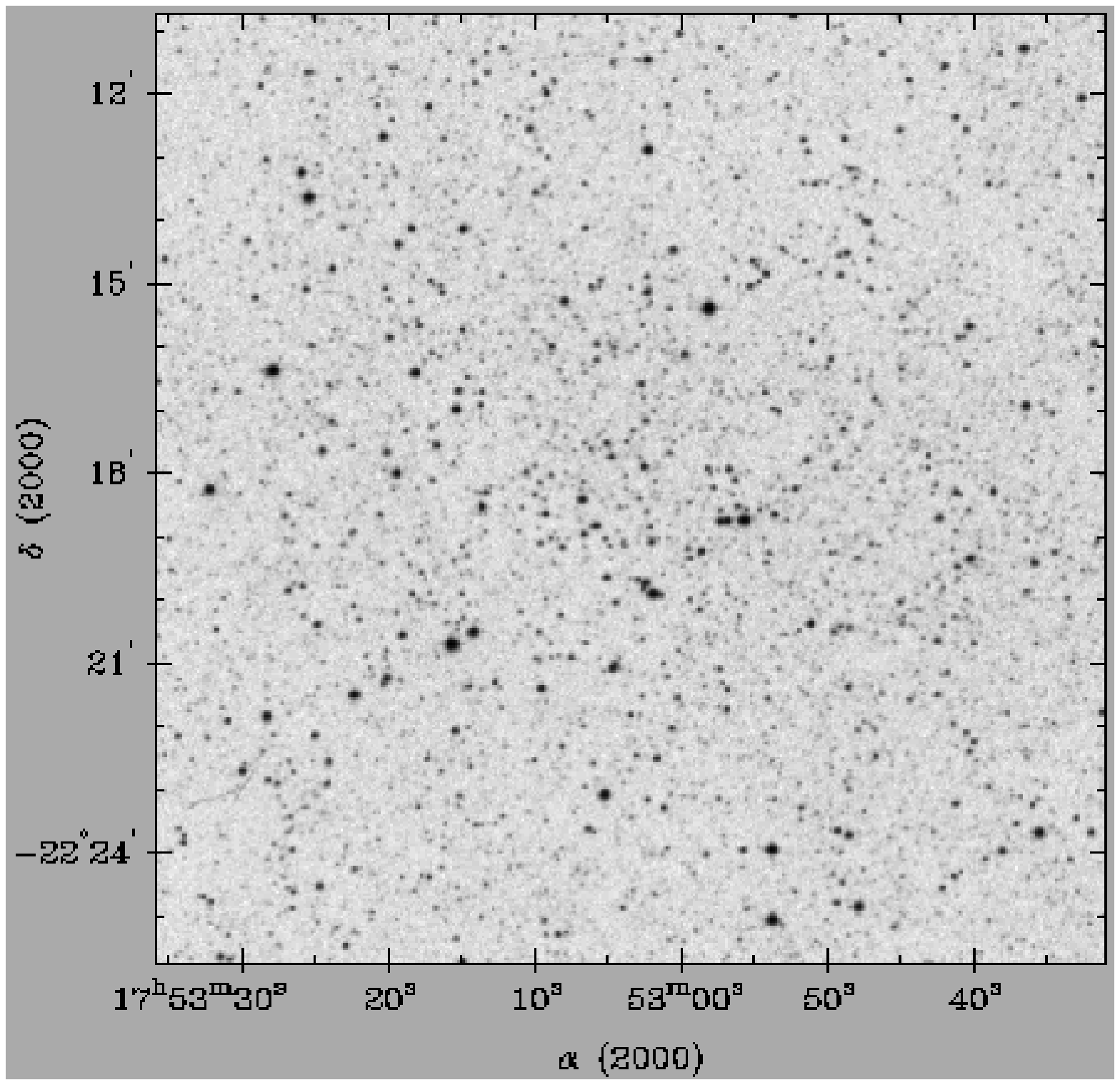}
\end{minipage}\hfill
\begin{minipage}[b]{0.50\linewidth}
\includegraphics[width=\textwidth]{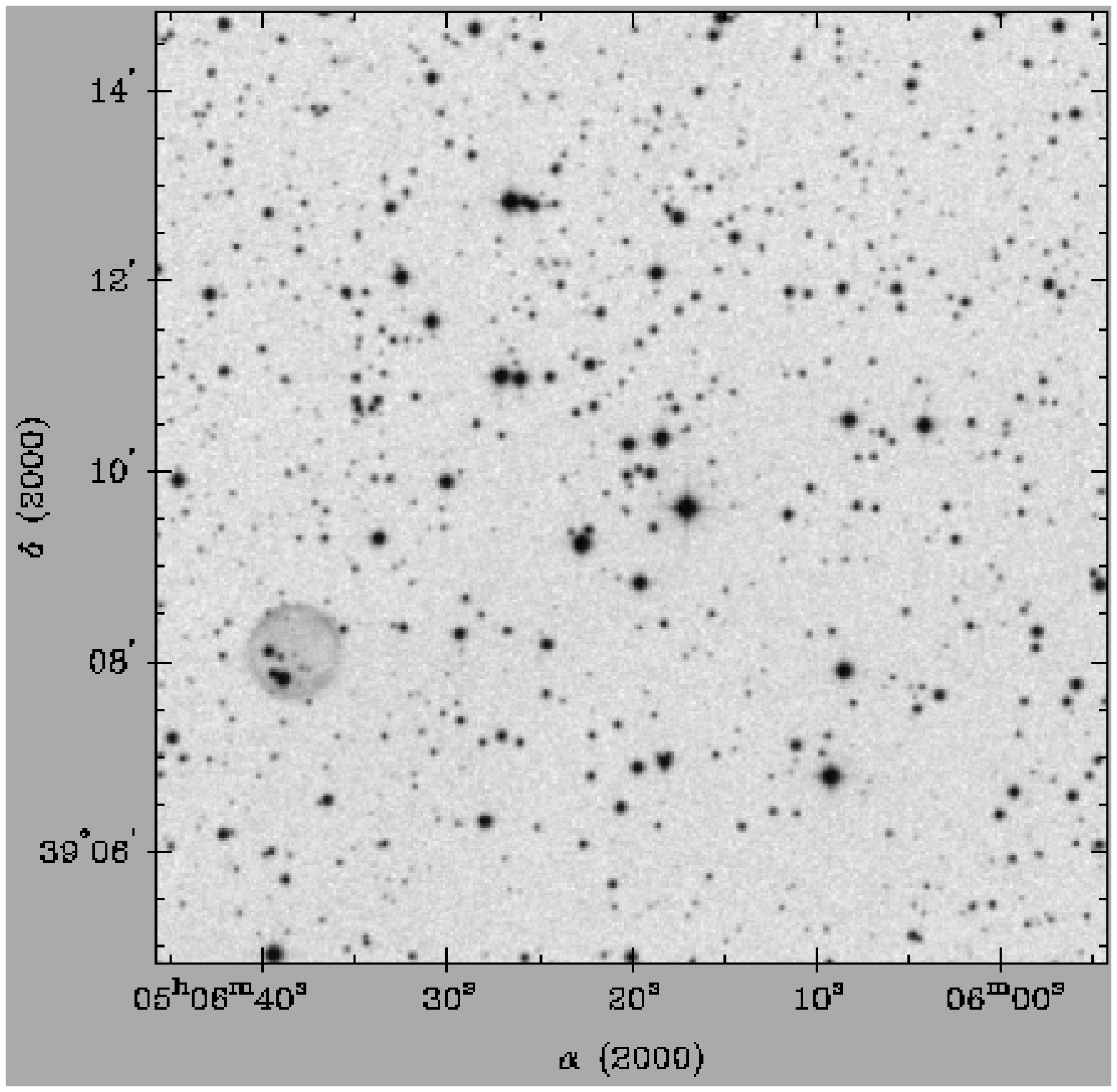}
\end{minipage}\hfill
\caption[]{Left panel: $15\arcmin\times15\arcmin$ DSS B image of the OC NGC\,6469 and the PN PK\,6+2.5.
The nearly stellar PN is located close to the lower-right corner (see Equatorial coordinates in 
Table~\ref{tab2}). Right panel: $10\arcmin\times10\arcmin$ XDSS R image of the OC New\,Cluster\,1 and the 
PN PK\,167-0.1.}
\label{fig2}
\end{figure*}

\begin{table*}
\caption[]{General data on the clusters}
\label{tab1}
\tiny
\renewcommand{\tabcolsep}{0.7mm}
\renewcommand{\arraystretch}{1.5}
\begin{tabular}{lcccccccccccccc}
\hline\hline
&\multicolumn{2}{c}{WEBDA}&&\multicolumn{10}{c}{Derived from 2MASS}\\
\cline{2-3}\cline{5-13}
Cluster&$\alpha(2000)$&$\delta(2000)$&&\rx&$\alpha(2000)$&$\delta(2000)$&$\ell$&$b$&Age&$\ebv$&\ds&\dgc&Alternative names\\
&(hms)&($^\circ\,\arcmin\,\arcsec$)&&(\arcmin)&(hms)&($^\circ\,\arcmin\,\arcsec$)&($^\circ$)&($^\circ$)&(Myr)&&(kpc)&(kpc)&&\\
(1)&(2)&(3)&&(4)&(5)&(6)&(7)&(8)&(9)&(10)&(11)&(12)&(13)\\
\hline
NGC\,2818A&09:16:01&$-$36:37:30&&80&09:16:07.7&$-$36:38:09.6&262.0&$+8.6$&$1000\pm100$&$0.10\pm0.03$&$2.8\pm0.1$
      &$8.1\pm0.2$ &ESO\,372$-$13,Hen\,2$-$23,PK\,261$+$08\\
M\,46&07:41:46&$-$14:48:36&&80&07:41:41.1&$-$14:51:45.0&231.9&$+4.0$&$250\pm50$&$0.10\pm0.03$
          &$1.5\pm0.1$&$8.3\pm0.2$  &NGC\,2437,Mel$-$75,Cr\,159,OCl$-$601\\
NGC\,6469&17:53:13&$-$22:19:11&&45&17:53:03.4&$-$22:18:14.4&6.55&$+1.97$&$250\pm50$&$0.58\pm0.06$
           &$1.1\pm0.1$&$6.1\pm0.2$   &Mel$-$182,Cr\,353,OCl$-$21,ESO\,589\,SC\,18 \\
New\,Cluster\,1&---&---&&60&05:06:20&$+$39:09:50.0&167.0&$-1.0$&$1000\pm100$&$0.29\pm0.03$
        &$1.7\pm0.1$&$8.9\pm0.2$  & \\
\hline
\end{tabular}
\begin{list}{Table Notes.}
\item Cols.~2 and 3: coordinates from WEBDA. Col.~4: extraction radius. Col.~10: reddening in the
cluster's central region (Sect.~\ref{age}). Col.~12: \dgc\ for $\rs=7.2$\,kpc (\citealt{GCProp}).
\end{list}
\end{table*}

Besides the investigation on possible PN/OC associations, the present work also serves to
provide reliable astrophysical parameters of scarcely studied OCs, as well as to characterise
a new one. Such data are important both to studies of the disc structure and to constrain
theories of molecular cloud fragmentation, star formation, as well as stellar and dynamical
evolutions. The present work employs near-IR \jj, \hh, and \ks\ photometry obtained from the
2MASS\footnote{The Two Micron All Sky Survey, All Sky data release (\citealt{2mass1997}), available
at {\em http://www.ipac.caltech.edu/2mass/releases/allsky/}} Point Source Catalogue (PSC). The
spatial and photometric uniformity of 2MASS, which allow extraction of large surrounding fields
that provide high star-count statistics, have been important to derive cluster parameters and
probe the nature of stellar overdensities (e.g. \citealt{ProbFSR}).

To this purpose we have developed quantitative tools to disentangle cluster and field stars in CMDs, in
particular two different kinds of filters. Basically we apply {\em (i)} field-star decontamination to uncover
cluster evolutionary sequences from the field, which is important to derive reddening, age, and distance
from the Sun, and {\em (ii)} colour-magnitude filters, which proved to be essential for building
intrinsic stellar radial density profiles (RDPs), as well as luminosity and mass functions. In particular,
field-star decontamination constrains more the age and distance, especially for low-latitude OCs
(\citealt{DiskProp}). These tools were applied to OCs and embedded clusters to enhance main-sequence (MS)
and/or pre-MS sequences with respect to the field (\citealt{M52N3960}; \citealt{N4755}; \citealt{N6611};
\citealt{5LowContr}; \citealt{M11}; \citealt{DetAnalOCs}). They were useful in the analysis of faint and/or distant
OCs (\citealt{CygOB2}; \citealt{3OpticalCl}; \citealt{5LowContr}; \citealt{FaintOCs}). In addition, more
constrained structural parameters, such as core  and limiting radii (\rc\ and \rl, respectively), and mass
function (MF) slopes, have been derived from colour-magnitude-filtered photometry (e.g. \citealt{DetAnalOCs}).

This paper is organised as follows. Sect.~\ref{Target_OCs} contains basic properties and reviews
literature data (when available) on the possible PN/OC associations. In Sect.~\ref{2mass} we present the
2MASS photometry, build CMDs, discuss the field-star decontamination, and derive cluster fundamental
parameters. Sect.~\ref{struc} describes cluster structure by means of stellar RDPs and mass density
profiles (MDPs). In Sect.~\ref{MF} MFs are built and cluster masses are estimated. In Sect.~\ref{CWODS}
aspects related to the structure of the present objects are discussed. In Sect.~\ref{Discus} we discuss
membership considerations. Concluding remarks are given in Sect.~\ref{Conclu}.

\section{The present possible PN/OC associations}
\label{Target_OCs}

In Fig.~\ref{fig1} we show optical XDSS\footnote{Extracted from the Canadian Astronomy Data Centre
(CADC), at \em http://cadcwww.dao.nrc.ca/} images of NGC\,2818/NGC\,2818A (left panel, R band) and
NGC\,2438/M\,46 (right panel, DSS B band). Fig.~\ref{fig2} contains a DSS B image of PK\,6+2.5/NGC\,6469
(left panel) and an XDSS R image of PK\,167-0.1/New\,Cluster\,1 (right panel).

Table~\ref{tab1} provides fundamental data on the OCs. WEBDA\footnote{\em obswww.univie.ac.at/webda -
\citet{Merm03}} coordinates are in cols.~2 and 3; col.~4
gives the 2MASS extraction radii (Sect.~\ref{2mass}). However, since the RDPs (Sect.~\ref{struc}) built
with the 2MASS coordinates presented a dip at the centre, new coordinates were searched to maximise the
central density of stars. The optimised central coordinates and the corresponding
Galactic longitude and latitude are given in Cols.~5 to 8 of Table~\ref{tab1}. Age, central reddening,
distance from the Sun and Galactocentric distance based on 2MASS data (Sect.~\ref{age}) are given in
Cols.~9 to 12. Additional designations are in col.~13.

Positional data of the PNe are given in Table~\ref{tab2}, where we also include the angular
separation with respect to the OC centre and its relation with the OC's core and limiting
radii (Sect.~\ref{Discus}). Coordinates are taken from
SIMBAD\footnote{http://simbad.u-starsbg.fr/simbad/ }.

Additional information on the possible pairs, related to the present work, are summarised below.

\subsection{The pair NGC\,2818/NGC\,2818A}
\label{Pair1}

Since NGC\,2818 refers to the original description of the PN, we adopt NGC\,2818A as the OC 
designation. \citet{MTL07} classify this pair as a suspected physical association. \citet{Pedreros89} 
derived a distance from the Sun $\ds=2300$\,pc and a reddening $\ebv=0.18$, which are consistent 
with $\ds=2660\pm830$\,pc (\citealt{Zhang95}) and $\ebv=0.28\pm0.15$ (\citealt{TASK92}) for the 
planetary nebula. Although earlier studies pointed to a similar radial velocity between cluster stars 
and the PN, \citet{Merm01} give $\rm V_r=20.7\pm0.3\,km\,s^{-1}$ for 15 red giants of NGC\,2818A, 
while for the PN $\rm V_r=-0.9\pm2.9\,km\,s^{-1}$ (\citealt{DAZ98}) and $\rm V_r=-1\pm3\,km\,s^{-1}$
(\citealt{MWF88}). \citet{Merm01} favour a projection effect. As for NGC\,2818A, available age 
estimates are $1$\,Gyr (\citealt{Merm01}), $930$\,Myr (\citealt{Lata02}) and $794$\,Myr 
(\citealt{Tad2002}). Additionally, \citet{Tad2002} provide for NGC\,2818A $\ds=2.9$\,kpc, cluster 
mass $m\approx288\,\ms$ and the limiting radius $\rl=3.9$\,pc. {\bf \citet{Merm01} and 
\citet{Lata02} derived $\ebv=0.18$, in agreement with the $\ebv=0.20$ of \citet{Tad2002}. }

\subsection{The pair NGC\,2438/M\,46 (NGC\,2437)}
\label{Pair2}

Based on the available evidence, \citet{MTL07} do not exclude the physical association. The distance 
from the Sun and reddening of the PN are $\ds=1775\pm630$\,pc (\citealt{Zhang95}) and $\ebv=0.17\pm0.08$
(\citealt{TASK92}), respectively. The isochrone fit with 2MASS photometry by \citet{MTL07} provided 
for the OC M\,46 $\ds=1700\pm250$\,pc, $\ebv=0.13\pm0.05$, and $\rm age=22\times10^7$\,Myr. Early 
determinations of radial velocity indicated differences of $\rm\approx30\,km\,s^{-1}$ (\citealt{MTL07},
and references therein), which suggest spatial coincidence. However, the most recent study on radial 
velocity of cluster stars and planetary nebula (\citealt{PK96}) shows comparable values, reinforcing 
the possible physical association. Additional available data on the OC M\,46 are $\rm age=250$\,Myr,
$\ds=1.5$\,kpc, core $\rc=2.27\pm0.13$\,pc, and limiting radii $\rl=11.6$\,pc (\citealt{Nilakshi2002});
$\ebv=0.15$, $\rm age=144$\,Myr, $\ds=1.4$\,kpc, $\rc=7.2\arcmin$, and $\rl=23\arcmin$ 
(\citealt{Kharchenko05}); $\rc=4.2\pm0.3$\,pc and $\rl=11$\,pc (\citealt{Sharma06}).

\subsection{The pair PK\,6+2.5/NGC\,6469}
\label{Pair3}

The PN is also known as M\,1-31, Ve\,3-59, ESO\,589-PN\,16, and PNG\,006.4+02.0. No distance to the PN
is available, but \citet{DAZ98} provide $\rm V_r=68.8\pm1.8\,km\,s^{-1}$. The PN is projected on the
bulge, which decreases the possibility of a physical association. \citet{Kharchenko05} obtained for 
the OC NGC\,6469 $\rm age=230$\,Myr, $\ebv=0.30$, $\ds=0.55$\,kpc, $\rc=5.4\arcmin$, and
$\rl=9\arcmin$.

\subsection{The pair PK\,167-0.1/New\,Cluster\,1}
\label{Pair4}

The PN is also known as A\,55 7 and PNG\,167.0-00.9. The available distance and radial velocity
estimates are $\ds=1780$\,pc (\citealt{Phillips04}) and $\rm V_r=58.2\pm6.5\,km\,s^{-1}$
(\citealt{DAZ98}), respectively. The OC New\,Cluster\,1 was found by one of us (E.B.) on
Digitised Sky Survey images, and is analysed for the first time.

\section{2MASS photometry}
\label{2mass}

\jj, \hh, and \ks\ 2MASS photometry was extracted in circular fields centred on the optimised coordinates
of the objects (cols.~5 and 6 of Table~\ref{tab1}) using VizieR\footnote{\em
http://vizier.u-strasbg.fr/viz-bin/VizieR?-source=II/246}. Previous analyses of OCs in different environments
(Sect.~\ref{Intro}) have shown that as long as no other populous cluster is present in the field, and
differential absorption is not prohibitive, wide extraction areas provide the required statistics for a
consistent field-star characterisation in terms of colour and luminosity distribution. Thus, the 2MASS extraction
radii (col.~4 of Table~\ref{tab1}) are significantly larger than the limiting radii (Sect.~\ref{struc} and
col.~7 of Table~\ref{tab4}) of the present objects. For decontamination purposes, comparison fields were
selected within wide rings centred on the cluster coordinates and beyond their limiting radii.
As a photometric quality constraint, 2MASS extractions
were restricted to stars with magnitudes {\em (i)} brighter than those of the 99.9\% Point Source Catalogue
completeness limit\footnote{Following the Level\,1 Requirement, according to {\em
http://www.ipac.caltech.edu/2mass/releases/allsky/doc/sec6\_5a1.html }} in the cluster direction, and
{\em (ii)} with errors in \jj, \hh, and \ks\ smaller than 0.2\,mag. The 99.9\% completeness limits are
different for each cluster, varying with Galactic coordinates. The fraction of stars with \jj, \hh, and \ks\
uncertainties smaller than 0.06\,mag is $\approx75\%$ (NGC\,6469), $\approx94\%$ (New\,Cluster\,1), 
$\approx73\%$ (NGC\,2818A), and $\approx93\%$ (M\,46). {\bf A typical distribution of uncertainties as 
a function of magnitude, for objects projected towards the central parts of the Galaxy, can be found 
in \citet{BB07}. } Reddening transformations use the relations
$A_J/A_V=0.276$, $A_H/A_V=0.176$, $A_{K_S}/A_V=0.118$, and $A_J=2.76\times\ejh$ (\citealt{DSB2002}), for
a constant total-to-selective absorption ratio $R_V=3.1$.

\begin{table}
\caption[]{PN coordinates and angular separation with the OCs}
\label{tab2}
\renewcommand{\tabcolsep}{0.7mm}
\renewcommand{\arraystretch}{1.25}
\begin{tabular}{cccccccccc}
\hline\hline
PN&$\alpha(2000)$&$\delta(2000)$&$\ell$&$b$&&$\Delta\,R$&$f_c$&$f_l$ \\
  &(hms)&($^\circ\,\arcmin\,\arcsec$)&($^\circ$)&($^\circ$)&&(\arcmin)&&\\
(1)&(2)&(3)&(4)&(5)&&(6)&(7)&(8)\\
\hline
NGC\,2818&09:16:01.7&$-$36:37:38.8&261.98&$+8.58$&&1.6&$0.8$&$0.1$\\
NGC\,2438&07:41:51.4&$-$14:43:54.9&231.80&$+4.12$&&8.2&$1.8$&$0.3$\\
PK\,6+2.5&17:52:41.4&$-$22:21:56.8&6.45&$+2.01$&&6.6&$6.0$&$0.5$\\
PK\,167-0.1&05:06:38.4&$+$39:08:8.6&167.04&$-0.97$&&4.9&$10.5$&$1.4$\\
\hline
\end{tabular}
\begin{list}{Table Notes.}
\item Cols.~2-5: coordinates from SIMBAD.
Col.~6: angular separation of the PN to the optimised cluster centre. Cols.~7 and 8: ratio
of the PN's angular separation with the core ($f_c=\Delta\,R/\rc$) and limiting
($f_l=\Delta\,R/\rl$) radii (Sect. ~\ref{struc}), respectively.
\end{list}
\end{table}

\begin{figure*}
\begin{minipage}[b]{0.50\linewidth}
\includegraphics[width=\textwidth]{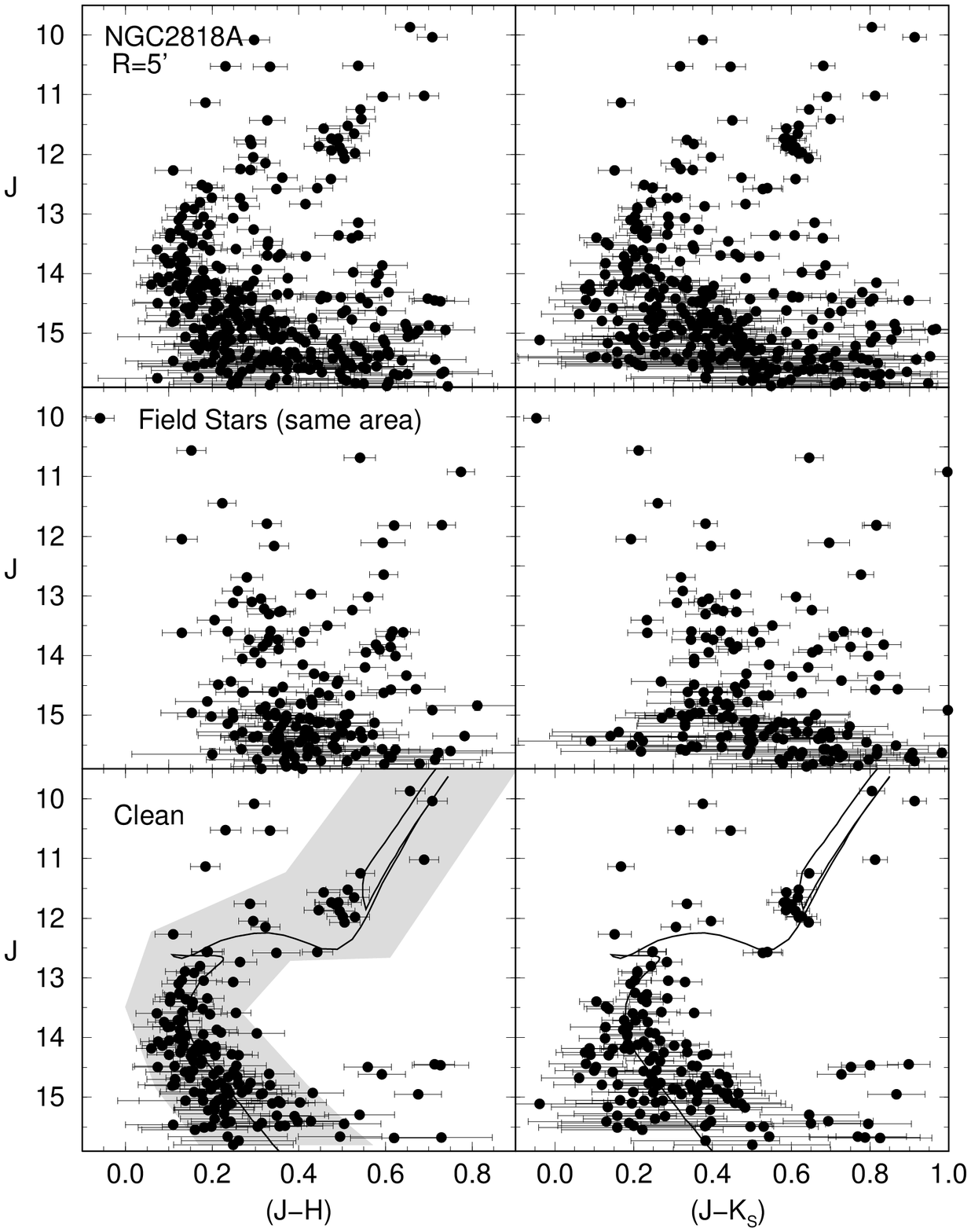}
\end{minipage}\hfill
\begin{minipage}[b]{0.50\linewidth}
\includegraphics[width=\textwidth]{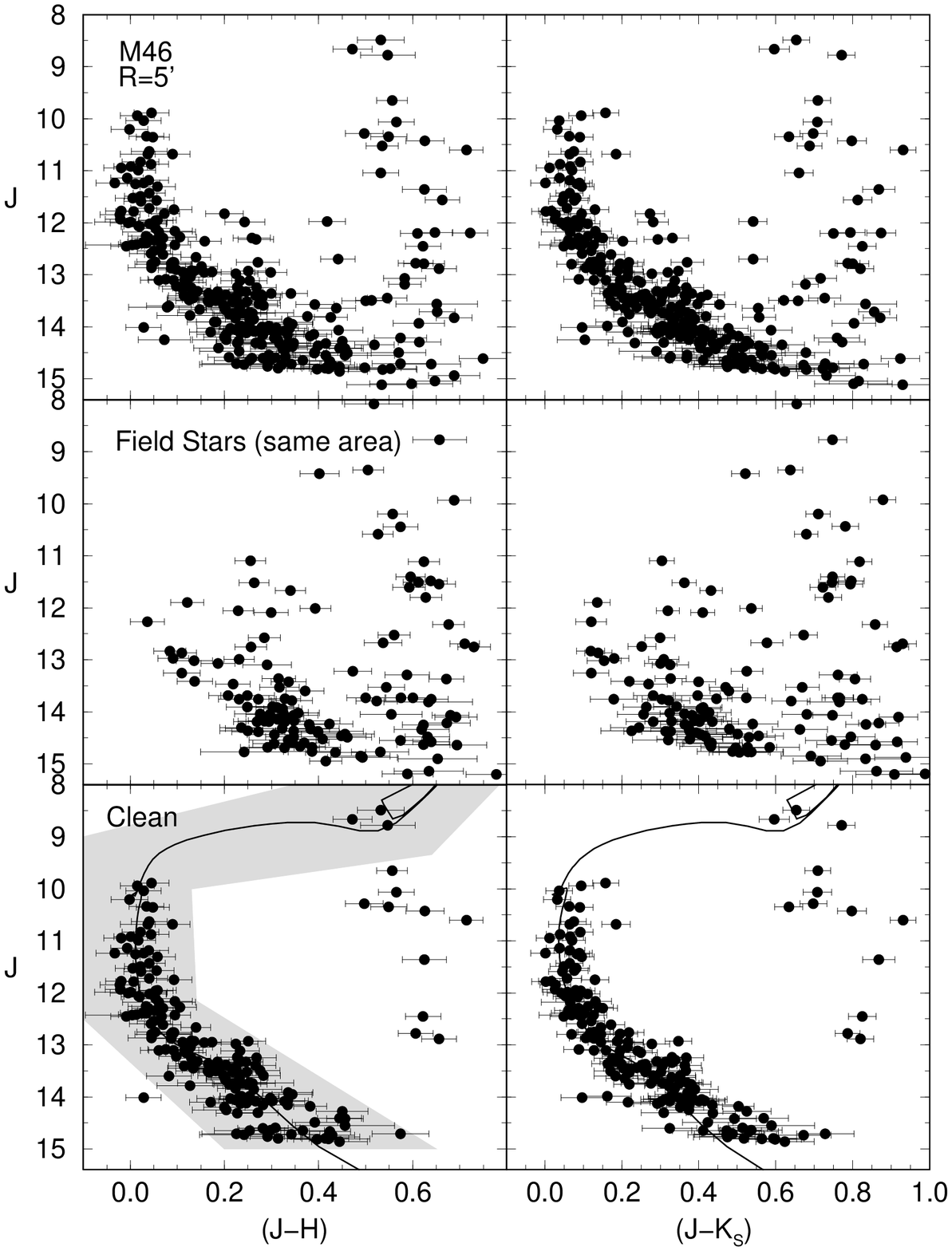}
\end{minipage}\hfill
\caption[]{Left: 2MASS CMDs extracted from the $R<5\arcmin$ region of NGC\,2818A. Top panels: 
observed photometry with the colours $\jj\times\jh$ (left) and $\jj\times\jk$ (right). Middle:
equal-area extraction of the comparison field. Besides some contamination of disk stars, a
populous MS and a  giant clump show up. Bottom panels: decontaminated CMDs set with the 1\,Gyr
Padova isochrone (solid line). The colour-magnitude filter used to isolate cluster MS/evolved
stars is shown as a shaded region. Right: Same for M\,46, except for the 250\,Myr Padova 
isochrone.}
\label{fig3}
\end{figure*}

\begin{figure*}
\begin{minipage}[b]{0.50\linewidth}
\includegraphics[width=\textwidth]{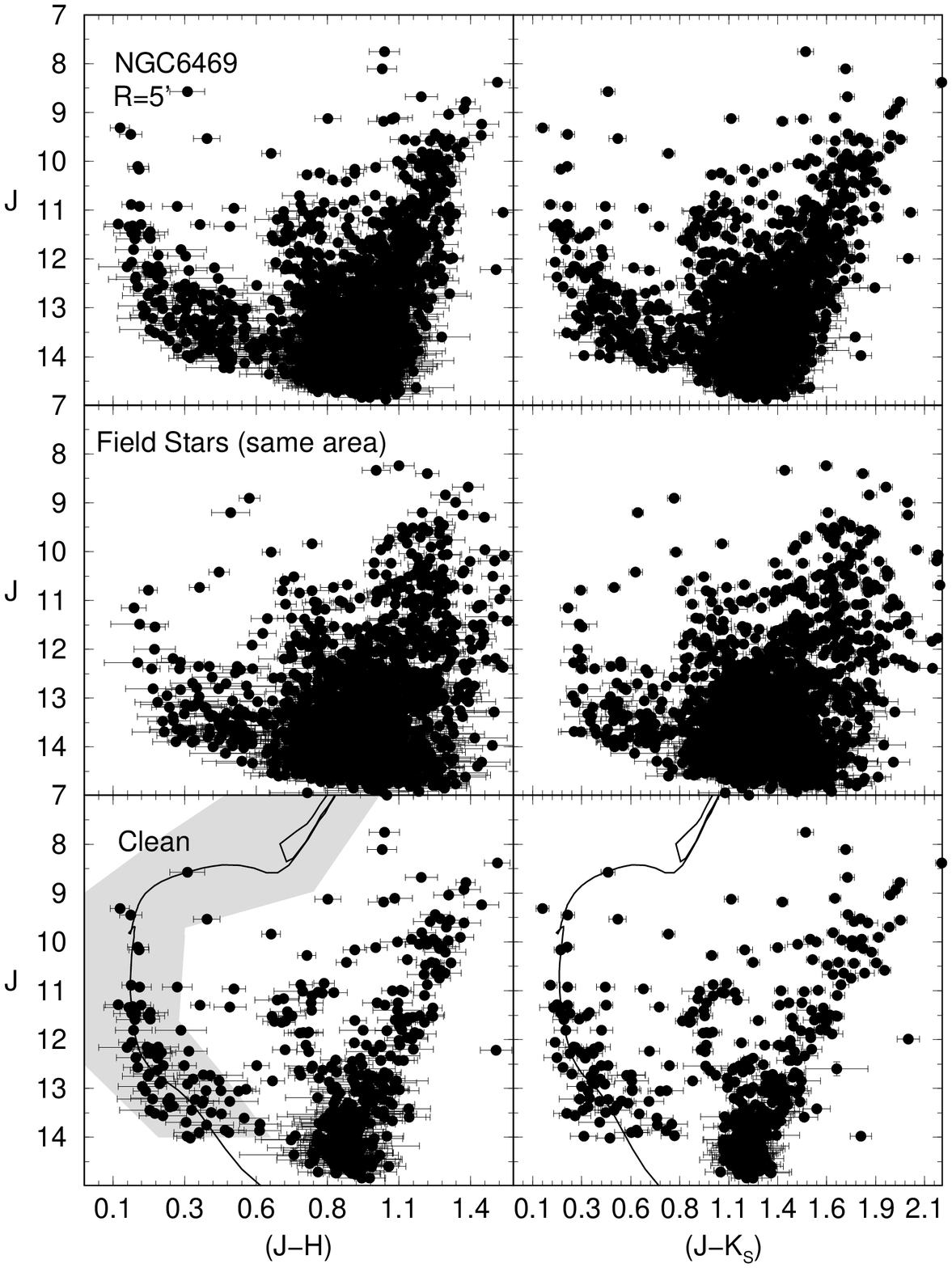}
\end{minipage}\hfill
\begin{minipage}[b]{0.50\linewidth}
\includegraphics[width=\textwidth]{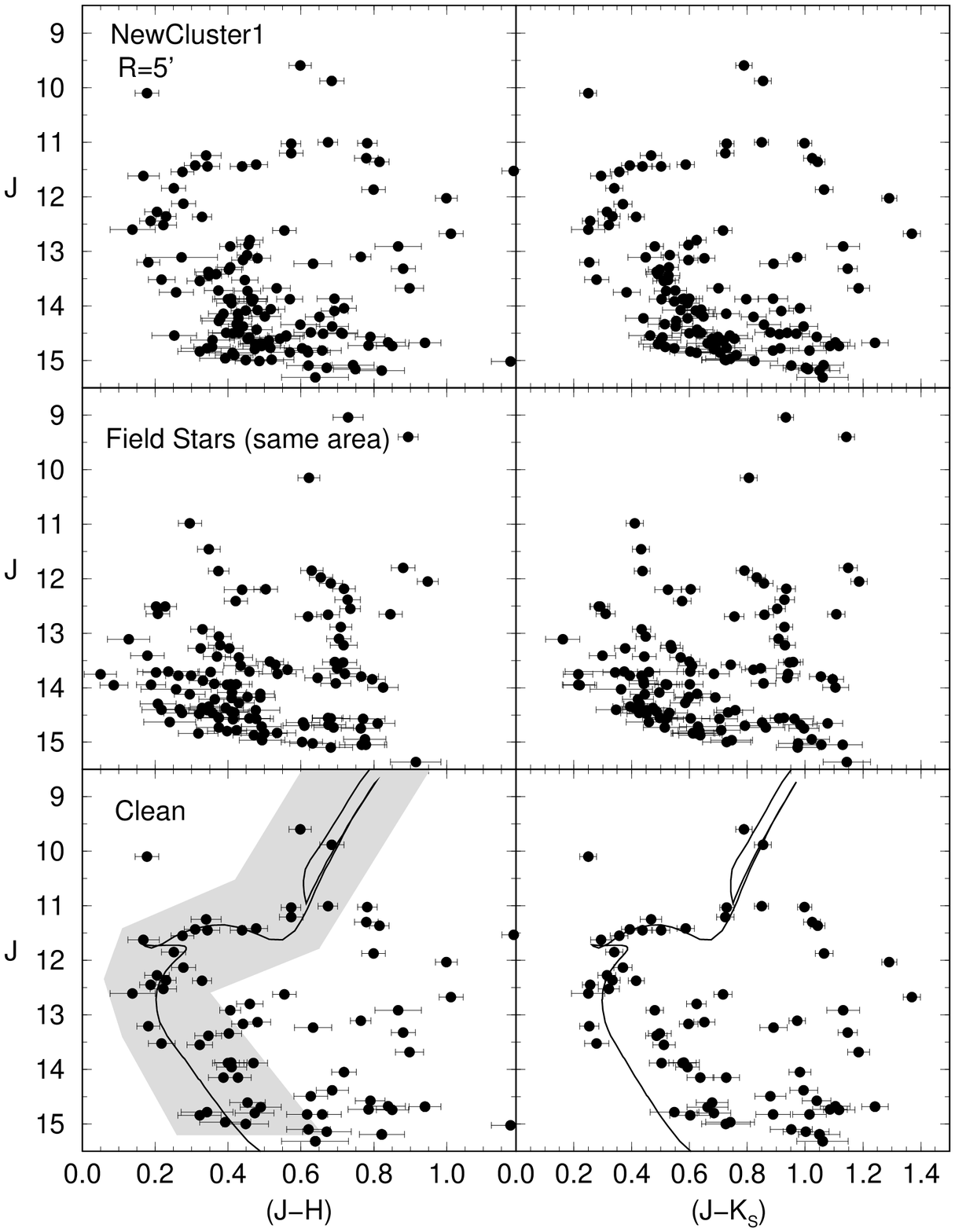}
\end{minipage}\hfill
\caption[]{Same as Fig.~\ref{fig3} for NGC\,6469 (left) and New\,Cluster\,1 (right). Isochrones
used are the 250\,Myr (NGC\,6469) and 1\,Gyr (New\,Cluster\,1) ones. Despite the heavy contamination by
bulge stars, a blue sequence shows up in the CMD of NGC\,6469. New\,Cluster\,1 is a poorly-populated
IAC.}
\label{fig4}
\end{figure*}

\subsection{Field-star decontaminated CMDs}
\label{Decont_CMDs}

$\jj\times\jh$ and $\jj\times\jk$ CMDs of central extractions of the clusters are shown in Figs.~\ref{fig3}
and \ref{fig4}. Features present in the central CMDs and those in the respective comparison field (top and
middle panels), show that field stars contribute in varying proportions to the CMDs, especially for the
bulge-projected OC NGC\,6469 (Fig.~\ref{fig4}). Nevertheless, when compared to the equal-area comparison
field extractions, cluster-like sequences are suggested, especially for M\,46, NGC\,2818A, and NGC\,6469
(the blue sequence).

To objectively quantify the field-star contamination in the CMDs we apply the statistical algorithm described
in \citet{BB07}. It measures the relative number densities of probable field and cluster stars in cubic CMD
cells whose axes correspond to the magnitude \jj\ and the colours \jh\ and \jk. These are the 2MASS colours
that provide the maximum variance among CMD sequences for OCs of different ages (e.g. \citealt{TheoretIsoc}).
The algorithm {\em (i)} divides the full range of CMD magnitude and colours into a 3D grid, {\em (ii)} computes
the expected number density of field stars in each cell based on the number of comparison field stars with
similar magnitude and colours as those in the cell, and {\em (iii)} subtracts the expected number of field
stars from each cell. By construction, the algorithm is sensitive to local variations of field-star
contamination with colour and magnitude (\citealt{BB07}). Typical cell dimensions are $\Delta\jj=0.5$, and
$\Delta\jh=\Delta\jk=0.25$, which are large enough to allow sufficient star-count statistics in individual
cells and small enough to preserve the morphology of different CMD evolutionary sequences. As comparison
field we use wide rings extracted from the region $\rl<R\la\rx$ around the cluster centre to obtain 
representative background star-count statistics, where \rl\ is the limiting radius (Sect.~\ref{struc}). 
{\bf The comparison fields effectively used are located within $\rm R=30\arcmin - 80\arcmin$ (NGC\,2818A), 
$\rm R=40\arcmin - 80\arcmin$ (M\,46), $\rm R=20\arcmin - 45\arcmin$ (NGC\,6469), and 
$\rm R=10\arcmin - 60\arcmin$ (New\,Cluster\,1). In all cases, the inner boundary of the comparison
field lies beyond the probable tidal radius (Sect.~\ref{struc}).} We emphasise that the equal-area
extractions shown in the middle panels of Figs.~\ref{fig3} and \ref{fig4} serve only for visual comparison
purposes. Actually, the decontamination process is carried out with the large surrounding area as
described above. Further details on the algorithm, including discussions on subtraction efficiency
and limitations, are given in \citet{BB07}.

The decontaminated CMDs are shown in the bottom panels of Figs.~\ref{fig3} and \ref{fig4}. As expected,
most of the disk contamination in NGC\,2818A, M\,46 and New\,Cluster\,1 is removed, leaving stellar sequences
typical of evolved/intermediate-age OCs. The centrally-projected NGC\,6469, on the other hand, is so heavily
contaminated by bulge stars that the algorithm does not subtract all of the field stars. The residual bulge 
component is probably due to differential reddening in the field, since the expected Poisson residuals should 
be smaller. In fact, the residual bulge component (bottom panel) contains 398 stars, and the observed (top panel)
contains 1893. Since the cluster
sequence is significantly bluer than the bulge stars, both sequences can be unambiguously separated.
We note that in all cases, essentially the same CMD features show up in both colours. New\,Cluster\,1
has evidence of MS-depletion (Fig.~\ref{fig4}), suggesting advanced dynamical evolution.

NGC\,2818A and M\,46 have easy-to-detect cluster CMD sequences and, to some extent, the same applies to
NGC\,6469. However, the non-populous nature of New\,Cluster\,1 requires additional statistical evidence.
To this end we present in Table~\ref{tab3} the full statistics of the decontaminated sequences and
field stars, by magnitude bins. As a caveat we note that it is more meaningful to work with isolated
cluster sequences instead of the full photometric sample. The bulge residual in the CMD of NGC\,6469,
for instance, contains more stars per magnitude bin than the assumed cluster sequence, which would mask
the decontamination statistics. In this sense, we first isolate the cluster sequences by means of
appropriate colour-magnitude filters, which are used to exclude stars with colours different from
those of the assumed cluster sequence. They are wide enough to accommodate cluster MS and evolved star
colour distributions, allowing for the $1\sigma$ photometric uncertainties. Colour-magnitude filter widths
should also account for formation or dynamical evolution-related effects, such as enhanced fractions of
binaries (and other multiple systems) towards the central parts of clusters, since such systems tend to
widen the MS (e.g. \citealt{BB07}; \citealt{N188}; \citealt{HT98}; \citealt{Kerber02}). The colour-magnitude
filters for the present objects are shown in the bottom panels of Figs.~\ref{fig3} and \ref{fig4}.

Statistically relevant parameters to characterise the nature of a star cluster are: {\em (i)} \ns\ which,
for a given magnitude bin, corresponds to the ratio of the decontaminated number of stars to the $\rm1\sigma$
Poisson fluctuation of the number of observed stars, {\em (ii)} \sFS, which is related to the probability
that the decontaminated stars result from the normal star count fluctuation in the comparison field
and, {\em (iii)} \fsU, which measures the star-count uniformity of the comparison field. See below for
the precise definition of these parameters. Properties of \ns,
\sFS, and \fsU, measured in OCs and field fluctuations are discussed in \citet{ProbFSR}. Table~\ref{tab3}
also provides integrated values of the above parameters, which correspond to the full magnitude range spanned
by each OC. The spatial regions considered here are those sampled by the CMDs shown in the top panels of
Figs.~\ref{fig3} and \ref{fig4}.

CMDs of star clusters should have integrated \ns\ values significantly larger than 1 (\citealt{ProbFSR}).
Indeed, this condition is met by NGC\,2818A, M\,46, and NGC\,6469. New\,Cluster\,1, on the other hand,
occurs at the $\approx2.7\sigma$ level, close to the lower limit observed in central directions
(\citealt{ProbFSR}). Values higher than $3\sigma$ occur as well for the \ns\ measured in magnitude bins.
Because of the small number of bright stars, we point out that this analysis should be basically
considered for $\jj\ga10$. As a further test of the statistical significance of the above results we
investigate star count properties of the field stars. First, the comparison field is divided in 8
sectors, each with $45^\circ$ of opening angle around the cluster centre. Next, we compute the parameter
\sFS, which is the $\rm 1\,\sigma$ Poisson fluctuation around the mean of the star counts
measured in the 8 sectors of the comparison field (corrected for the different areas of the sectors and
cluster extraction). For a spatially uniform comparison field, \sFS\ should be very small. In this context,
star clusters should have the probable number of member stars (\nc) higher than $\sim3\,\sFS$, to minimise
the probability that \nc\ results from the normal fluctuation of a non-uniform comparison field. Again,
this condition is fully satisfied, in some cases reaching the level $\nc\sim10\,\sFS$. The ratio decreases
somewhat for New\,Cluster\,1, but still is higher than $\sim3$, probably because it is almost projected
against the anti-centre. Finally, we also provide in Table~\ref{tab3} the parameter \fsU. For a
given magnitude bin we first compute the average number of stars over all sectors $\langle N\rangle$
and the corresponding $\rm 1\sigma$ fluctuation $\sigma_{\langle N\rangle}$; thus, \fsU\ is defined
as $\rm\fsU=\sigma_{\langle N\rangle}/\langle N\rangle$. Non uniformities such as heavy differential
reddening should result in high values of \fsU.

\begin{table*}
\caption[]{Statistics of the field-star decontamination in magnitude bins}
\label{tab3}
\renewcommand{\tabcolsep}{1.05mm}
\renewcommand{\arraystretch}{1.2}
\begin{tabular}{ccccccccccccccccccccccccc}
\hline\hline
\multicolumn{24}{c}{Colour-magnitude filtered photometry}\\
\cline{1-24}
&\multicolumn{5}{c}{NGC\,2818A}&&\multicolumn{5}{c}{M\,46}&&\multicolumn{5}{c}{NGC\,6469}&&
\multicolumn{5}{c}{New\,Cluster\,1}\\
\cline{2-6}\cline{8-12}\cline{14-18}\cline{20-24}
$\Delta\jj$&\sFS&\no&\nc&\ns&\fsU& &\sFS&\no&\nc&\ns&\fsU& &\sFS&\no&\nc&\ns&\fsU& &\sFS&\no&\nc&\ns&\fsU\\
\cline{1-24}
  8-9&---&---&---  &--- &---     &&0.11&3&3&1.7&0.28   &&0.14&1&1&1.0&0.37  &&---&---&---&---&--- \\
 9-10&0.14&1&1&1.0&0.22    &&0.25&2&2&1.4&0.39   &&0.31&3&3&1.7&0.45  &&0.29&2&2&1.4&0.45 \\
10-11&0.34&2&2&1.4&0.10  &&0.30&12&12&3.5&0.53&&0.74&5&5&2,2&0.46  &&---&---&---&---&--- \\
11-12&0.34&18&17&4.0&0.07&&0.31&22&22&4.5&0.26 &&1.92&14&12&3.2&0.40&&0.77&11&11&3.3&0.17 \\
12-13&1.36&20&12&2.7&0.14&&0.28&43&43&6.3&0.06 &&2.83&36&25&4.2&0.16&&1.07&7&5&1.9&0.15 \\
13-14&1.43&31&27&4.8&0.25&&1.16&72&72&6.4&0.05 &&7.01&67&27&3.3&0.14&&2.09&19&3&0.7&0.11 \\
14-15&3.47&78&63&7.1&0.18&&3.18&85&85&5.4&0.08 &&---&---&---&---&---    &&3.35&40&12&1.9&0.08 \\
15-16&4.51&81&36&4.0&0.09&&--- &---&---&---  &---  &&---&---&---&---&---    &&---&---&---&---&--- \\
\cline{2-6}\cline{8-12}\cline{14-18}\cline{20-24}
All &10.95&231&158&9.9&0.12&&4.97&238&182&11.6&0.07&&9.20&126&73&5.9&0.12&&6.42&79&33&2.7&0.09 \\
\hline
\end{tabular}
\begin{list}{Table Notes.}
\item This table provides, for each magnitude bin ($\Delta\jj$), the $\rm 1\,\sigma$ Poisson fluctuation
(\sFS) around the mean, with respect to the star counts measured in the 8 sectors of the comparison field,
the number of observed stars (\no) within the spatial region sampled in the CMDs shown in the top panels
of Figs.~\ref{fig3} and \ref{fig4}, the respective number of probable member stars (\nc) according to the
decontamination algorithm, the \ns\ parameter, and the field-star uniformity parameter. The statistical
significance of \nc\ is reflected in its ratio with the $1\sigma$ Poisson fluctuation of \no\ (\ns) and 
with \sFS. The bottom line corresponds to the full magnitude range.
\end{list}
\end{table*}

The three statistical tests applied to the present sample, i.e. {\em (i)} the decontamination algorithm,
{\em (ii)} the integrated and per magnitude \ns\ parameter, and {\em (iii)} the ratio of \nc\ to \sFS,
produce consistent results. As for New\,Cluster\,1, the number of decontaminated stars in each magnitude
bin is significantly larger than what could be expected from field-star fluctuations. Besides, its field
is rather uniform (Table~\ref{tab3}). We conclude that New\,Cluster\,1 is a poorly-populated Gyr-class OC.

\subsection{Cluster age, reddening, and distance}
\label{age}

Fundamental parameters for the present clusters are derived from CMD fits with solar-metallicity Padova
isochrones (\citealt{Girardi2002}) computed with the 2MASS \jj, \hh, and \ks\ filters\footnote{\em
http://stev.oapd.inaf.it/$\sim$lgirardi/cgi-bin/cmd}. These isochrones are very similar to the
Johnson-Kron-Cousins (e.g. \citealt{BesBret88}) ones, with differences of at most 0.01 in \jh\
(\citealt{TheoretIsoc}). The decontaminated CMD morphologies (bottom panels of Figs.~\ref{fig3} and
\ref{fig4}) provide enough constraints to derive reliable cluster fundamental parameters, which
are given in Table~\ref{tab1}. {\bf The isochrone fits to the decontaminated cluster CMDs are shown 
in the bottom panels of Figs.~\ref{fig3} and \ref{fig4}.}

{\bf From the isochrone fit to the CMD of NGC\,2818A (Fig.~\ref{fig3})} we derive the age 
$1.0\pm0.1$\,Gyr, reddening $\ejh=0.03\pm0.01$ that converts to $\ebv=0.10\pm0.02$ and 
$A_V=0.31\pm0.07$, and  $\ds=2.8\pm0.1$\,kpc. {\bf We adopt $\rs=7.2\pm0.3$\,kpc (\citealt{GCProp}) 
as the Sun's distance to the Galactic centre to compute the OC's Galactocentric distances. The latter 
value was derived by means of the globular cluster spatial distribution. Besides, other recent studies 
gave similar results, e.g. $\rs=7.2\pm0.9$\,kpc (\citealt{Eisen03}), $\rs=7.62\pm0.32$\,kpc
(\citealt{Eisen05}) and $\rs=7.52\pm0.10$\,kpc (\citealt{Nishiyama06}), with different approaches.
Thus, for} $\rs=7.2$\,kpc, the Galactocentric distance of
NGC\,2818A is $\dgc=8.1\pm0.2$\,kpc. {\bf NGC\,2818A lies $\approx0.9$\,kpc outside the Solar circle}. Within 
the uncertainties, the present age and reddening values agree with those of \citet{Lata02} and \citet{Merm01}. 
The distance from the Sun is the same as that in \citet{Tad2002}. The mass at the turnoff (TO) of NGC\,2818A is 
$\mTO=2.1\,\ms$.

Parameters of M\,46 are the age $250\pm50$\,Myr, $\ejh=0.03\pm0.01$, $\ebv=0.10\pm0.02$, and $A_V=0.31\pm0.07$,
$\ds=1.5\pm0.1$\,kpc, and $\dgc=8.3\pm0.2$\,kpc{\bf, $\approx1.1$\,kpc outside the Solar circle}. Within the
uncertainties, these parameters agree with those of \citet{MTL07} and \citet{Nilakshi2002}; the present age is 
about twice as that of \citet{Kharchenko05}. TO mass is $\mTO=3.5\,\ms$.

For the bulge-projected NGC\,6469 we derive the age $250\pm50$\,Myr, $\ejh=0.18\pm0.02$, $\ebv=0.58\pm0.05$
and $A_V=1.8\pm0.2$, $\ds=1.1\pm0.1$\,kpc, and $\dgc=6.1\pm0.2$\,kpc{\bf, $\approx1.1$\,kpc inside the Solar 
circle}. The present age agrees with that of
\citet{Kharchenko05}, but they found about twice the reddening and half the distance from the Sun.
TO mass is $\mTO=3.5\,\ms$.

Finally, parameters of New\,Cluster\,1, derived for the first time, are the age $1.0\pm0.1$\,Gyr,
$\ejh=0.09\pm0.01$, $\ebv=0.29\pm0.03$, and $A_V=0.9\pm0.1$, $\ds=1.7\pm0.1$\,kpc, and
$\dgc=8.9\pm0.2$\,kpc{\bf, $\approx1.7$\,kpc outside the Solar circle}. TO mass is $\mTO=2.1\,\ms$.

Except for the bulge-projected NGC\,6469, which is located $\approx1$\,kpc inside the Solar circle,
the remaining OCs are more than $1$\,kpc outside it.

\section{Cluster structure}
\label{struc}

Structural parameters are derived by means of RDPs, defined as the projected radial distribution of
the number density of stars around the cluster centre.

Star clusters usually have RDPs that follow some well-defined analytical profile.
The most often used are the single mass, modified isothermal sphere of \citet{King66}, the modified
isothermal sphere of \citet{Wilson75}, and the power law with a core of \citet{EFF87}. Each function
is characterised by different parameters that are somehow related to cluster structure. However,
because the error bars in the present RDPs are significant (see below), we decided to use the
analytical profile $\sigma(R)=\sigma_{bg}+\sigma_0/(1+(R/R_C)^2)$, where $\sigma_{bg}$ is
the residual background density, $\sigma_0$ is the central density of stars, and \rc\ is the core
radius. This function is similar to that introduced by \cite{King1962} to describe the surface
brightness profiles in the central parts of globular clusters.

\begin{table*}
\caption[]{Structural parameters of evolved $+$ MS stars (CM-filtered photometry)}
\label{tab4}
\renewcommand{\tabcolsep}{2.5mm}
\renewcommand{\arraystretch}{1.3}
\begin{tabular}{lcccccccccc}
\hline\hline
&&&\multicolumn{5}{c}{RDP}&&\multicolumn{2}{c}{MDP}\\
\cline{4-8}\cline{10-11}
Cluster&$1\arcmin$&&$\sigma_{bg}$&$\sigma_0$&$\delta_c$&\rc&\rl&&$\sigma_0$&\rc \\
       &(pc)&&$\rm(stars\,pc^{-2})$&$\rm(stars\,pc^{-2})$&&(pc)&(pc)&
       &$\rm(\ms\,pc^{-2})$&(pc)\\
(1)&(2)&&(3)&(4)&(5)&(6)&(7)&&(8)&(9)\\
\hline
NGC\,2818A&0.805&&$1.8\pm0.1$&$8.8\pm2.2$&$5.9\pm1.2$&$1.5\pm0.3$&$11\pm1$&&$13.4\pm1.8$&$1.5\pm0.2$\\
M\,46    &0.442&&$4.4\pm0.1$&$18.8\pm4.6$&$5.1\pm1.0$&$2.0\pm0.4$ &$12.4\pm0.8$&&$30.8\pm6.9$&$2.0\pm0.3$\\
NGC\,6469&0.318&&$9.8\pm0.2$&$35.3\pm13.0$&$4.6\pm0.5$&$0.35\pm0.10$ &$4.0\pm0.5$&&$49\pm8$&$0.45\pm0.10$\\
New\,Cluster\,1  &0.493&&$3.7\pm0.1$&$20.8\pm6.6$&$6.6\pm1.8$ &$0.23\pm0.06$&$1.7\pm0.2$&&$27.9\pm1.6$&$0.25\pm0.02$\\
\hline
\end{tabular}
\begin{list}{Table Notes.}
\item Col.~2: arcmin to parsec scale. King profile is expressed as
$\sigma(R)=\sigma_{bg}+\sigma_0/(1+(R/R_{\rm core})^2)$. To minimise degrees of freedom
in RDP fits, $\sigma_{bg}$ was kept fixed (measured in the respective comparison fields) while
$\sigma_0$ and \rc\ were allowed to vary. MDPs are background subtracted profiles. Col.~5:
cluster/background density contrast ($\delta_c=1+\sigma_0/\sigma_{bg}$), measured in CM-filtered
RDPs.
\end{list}
\end{table*}

In all cases we build the stellar RDPs with colour-magnitude filtered photometry (Sect.~\ref{Decont_CMDs}).
However, residual field stars with colours similar to those of the cluster are
expected to remain inside the colour-magnitude filter region. They affect the intrinsic stellar
RDP in a degree that depends on the relative densities of field and cluster
stars. The contribution of the residual contamination to the observed RDP is statistically taken
into account by means of the comparison field. In practical terms, the use of colour-magnitude
filters in cluster sequences enhances the contrast of the RDP with respect to the background level,
especially for objects in dense fields (e.g. \citealt{BB07}).

To avoid oversampling near the centre and undersampling at large radii, RDPs are built by counting
stars in rings of increasing width with distance {\bf from the centre. A typical distribution of
ring widths would be $\Delta\,R=0.5,\ 1,\ 2,\ 5,\ {\rm and}\ 10\arcmin$, respectively for 
$0\arcmin\le R<1\arcmin$, $1\arcmin\le R<4\arcmin$, $4\arcmin\le R<10\arcmin$, $10\arcmin\le R<30\arcmin$, 
and $R\ga30\arcmin$. However, t}he number and width of the rings are
adjusted to produce RDPs with adequate spatial resolution and as small as possible $1\sigma$ Poisson
errors. The residual background level of each RDP corresponds to the average number of colour-magnitude
filtered stars measured in the comparison field. The $R$ coordinate (and respective uncertainty) of each
ring corresponds to the average position and standard deviation of the stars inside the ring.

\begin{figure}
\resizebox{\hsize}{!}{\includegraphics{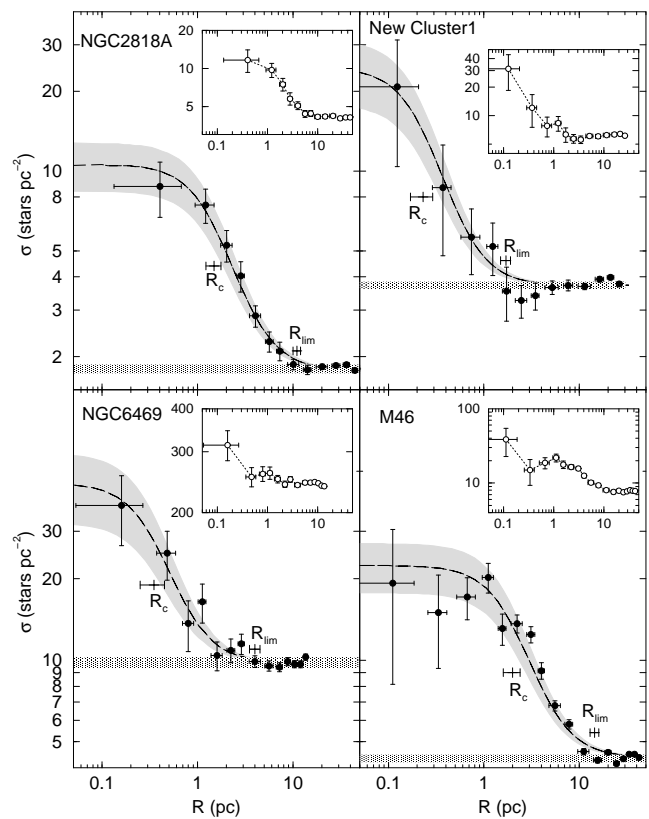}}
\caption[]{Stellar RDPs (filled circles) of the present star clusters built with colour-magnitude
filtered photometry.
Solid line: best-fit King-like profile. Horizontal shaded region: offset field stellar background level. Gray
regions: $1\sigma$ King fit uncertainty. The core and limiting radii are indicated. Insets: RDPs built with the 
observed photometry. Absolute scale is used.}
\label{fig5}
\end{figure}

The resulting RDPs of the present star clusters are given in Fig.~\ref{fig5}. For absolute comparison between
clusters the radius scale was converted to parsecs and the number density of stars to $\rm stars\,pc^{-2}$\
for the distances derived in Sect.~\ref{age}. Besides the RDPs resulting from the colour-magnitude filters, we
also show, for illustrative purposes, those produced with the observed (raw) photometry. In all cases,
minimisation of the number of non-cluster stars by the colour-magnitude filter resulted in RDPs with a
significantly higher contrast with respect to the background. Fits of the King-like profile were performed
with a non-linear least-squares fit routine that uses errors as weights. To minimise degrees of
freedom, $\sigma_0$ and \rc\ were derived from the RDP fit, while $\sigma_{bg}$ is measured in the respective
comparison field. These values are given in Table~\ref{tab4}, and the best-fit solutions are superimposed on
the colour-magnitude filtered RDPs (Fig.~\ref{fig5}). Because of the 2MASS photometric limit, which in most
cases corresponds to a cutoff for stars brighter than $\jj\approx15$, $\sigma_0$ should be taken as a lower
limit to the actual central number density. The adopted King-like function describes well the RDPs
throughout the full radii range, within uncertainties.

We also estimate the cluster limiting radius and uncertainty by visually comparing the RDP level (taking 
into account fluctuations) with the background. In this sense, \rl\ corresponds to the distance from the
cluster centre where RDP and background become statistically indistinguishable from each other (e.g.
\citealt{DetAnalOCs}, and references therein). For practical purposes, most of the cluster stars can be
considered to be contained within $\rl$. Note that \rl\ should not be mistaken for the tidal radius. For 
instance, in populous and relatively high Galactic latitude OCs such as M\,26, M\,67, NGC\,188, and 
NGC\,2477, limiting radii are a factor $\sim0.5 - 0.7$ of the respective tidal radii (\citealt{DetAnalOCs}). 
The limiting radii of the present objects are given in col.~7 of Table~\ref{tab4}. Tidal radii are derived 
from fits of King profile to RDPs, which depend on wide surrounding fields and adequate Poisson errors. 
{\bf If limiting and tidal radii of the present clusters are similarly related as for the bright ones, 
we note that, in all cases, the lower-limit of the radial range adopted as comparison field 
(Sect.~\ref{Decont_CMDs}) is located beyond the respective probable tidal radius. This, in turn, 
minimises the probability of cluster members at large radii, e.g. in the halo, to be considered as 
field stars by the decontamination algorithm.}

Table~\ref{tab4} (col.~5) also provides the density contrast parameter $\delta_c=1+\sigma_0/\sigma_{bg}$.
Since $\delta_c$ is measured in colour-magnitude-filtered RDPs, it does not necessarily correspond to the
visual contrast produced by observed stellar distributions in images (Figs.~\ref{fig1} and \ref{fig2}).
NGC\,6469, for instance, presents a very low contrast in the DSS B image (Fig.~\ref{fig2}) but, because
most of the non-cluster stars have been excluded by the colour-magnitude filter, the corresponding RDP
presents a relatively high density contrast, $\delta_c\approx4.6$. The same applies to the non-populous
OC New\,Cluster\,1.

\begin{figure}
\resizebox{\hsize}{!}{\includegraphics{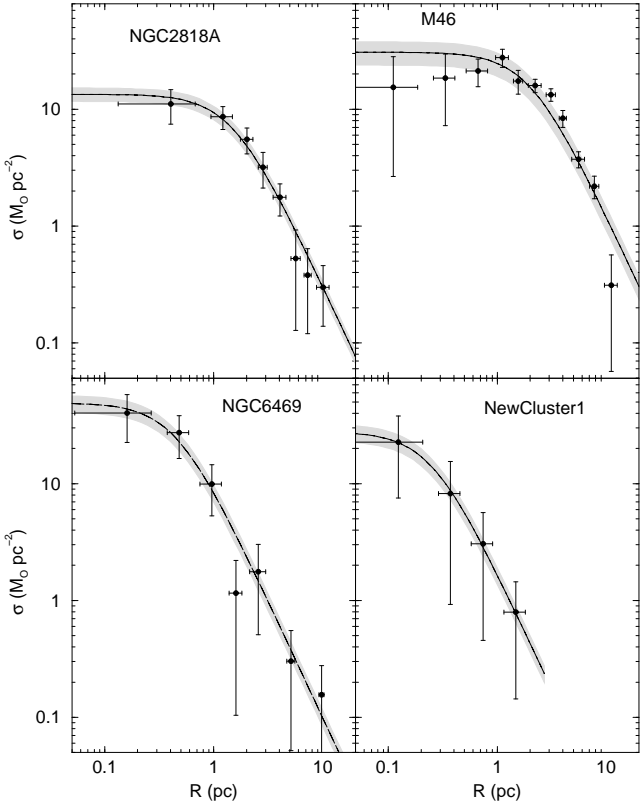}}
\caption[]{Same as Fig.~\ref{fig5} for the mass density profiles. The background contribution has
been subtracted.}
\label{fig6}
\end{figure}

Probably because of different methods and data sets, the present values of \rc\ and \rl\ are different
from those in common with \citet{Kharchenko05}. The difference, especially in \rc, may be attributed to
their brighter limits (\citealt{KPRSS04}) producing shallower profiles for the bulge-contaminated cluster
NGC\,6469.

\subsection{Mass Density Profiles}
\label{MDPs}

To complete the structural description of the objects we take the mass-luminosity (ML) relation derived
from isochrone fits (Sect.~\ref{age}) to build statistical mass-density profiles. We
follow the same systematics as that used to build RDPs. Instead of  computing the
number density of stars in rings, we now assign each star a mass according to the respective
ML relation. MDPs are produced by subtracting from the mass density in each ring that
measured in the comparison field. They are shown in Fig.~\ref{fig6}, together with the respective
King-like fits. Likewise RDPs, MDPs are well described by King-like profiles. Core radii derived
from MDPs (col.~9 of Table~\ref{tab4}) agree, at $1\,\sigma$, with RDP ones (col.~6).

\section{Mass functions and cluster mass}
\label{MF}

The methods presented in \citet{DetAnalOCs} (and references therein) are used to build the mass functions
(MFs), $\left(\phi(m)=\frac{dN}{dm}\right)$. We build them with colour-magnitude filtered photometry, the
three 2MASS bands separately, and the ML relations obtained from the respective Padova isochrones
and distances from the Sun (Sect.~\ref{age}). Further details on MF construction are given in \citet{FaintOCs}.
The effective magnitude range over which MFs are computed is that where clusters present an excess of stars
with respect to the comparison field. In all cases it begins right below the TO and ends at a faint magnitude
limit brighter than that stipulated by the 2MASS completeness limit (Sect.~\ref{2mass}). The effective MS stellar
mass ranges are $1.1\leq m(\ms)\leq2.1$ (NGC\,2818A), $1.1\leq m(\ms)\leq3.5$ (M\,46), $1.3\leq m(\ms)\leq3.5$
(NGC\,6469), and $1.1\leq m(\ms)\leq2.1$ (New\,Cluster\,1). However, we note that because of the non-populous
nature of New\,Cluster\,1, the MF error bars resulted exceedingly large.

\begin{table*}
\caption[]{Parameters related to cluster mass and stellar population}
\label{tab5}
\renewcommand{\tabcolsep}{2.4mm}
\renewcommand{\arraystretch}{1.25}
\begin{tabular}{cccccccccccc}
\hline\hline
&\multicolumn{2}{c}{Evolved}&&\multicolumn{3}{c}{Observed MS}&&
\multicolumn{4}{c}{$\rm Evolved\ +\ Extrapolated\ MS\ $}\\
\cline{2-3}\cline{5-7}\cline{9-12}
Region&$N^*$&$m$&&$\chi$&$N^*$&\mobs&&$N^*$&$m$&$\sigma$&$\rho$\\
&(stars)&(\ms)&&&(stars)&(\ms)&& ($10^2$stars)&($10^2\ms$)&($\rm \ms\,pc^{-2}$)&($\rm \ms\,pc^{-3}$)\\
 (1)  & (2)   & (3)  &&(4)     & (5)    & (6)   && (7)    & (8)   &(9) &(10) \\
\hline
&\multicolumn{11}{c}{NGC\,2818A --- MS: $1.1\leq m(\ms)\leq2.1$ --- $\rm{Age}=1.0\pm0.1$\,Gyr}\\
\cline{2-12}
Core&$6\pm1$&$14\pm3$&&---&$24\pm5$&$40\pm5$&&---&---&---&---\\
Halo&$18\pm8$ &$38\pm18$ &&---& $90\pm20$&$149\pm20$&&---&---&---&---\\
Overall& $25\pm9$&$52\pm18$ &&$1.44\pm0.55$& $115\pm20$&$189\pm21$
&&$41\pm36$&$15\pm8$&$3.9\pm2.0$&$0.26\pm0.14$\\
\cline{2-12}
&\multicolumn{11}{c}{M\,46 --- MS: $1.1\leq m(\ms)\leq3.5$ --- $\rm{Age}=250\pm50$\,Myr}\\
\cline{2-12}
Core&$9\pm2$&$29\pm6$&&$1.31\pm0.50$&$133\pm15$&$214\pm14$&&$17\pm13$&$7\pm3$&$55\pm21$&
     $21\pm8$\\
Halo&$22\pm6$ &$68\pm19$ &&$1.60\pm0.22$&  $727\pm70$&$1147\pm56$&&$112\pm83$&$43\pm16$&
     $9.0\pm3.3$&$0.54\pm0.20$\\
Overall& $32\pm6$&$98\pm20$ &&$1.63\pm0.15$& $860\pm71$&$1362\pm58$&&$135\pm99$&$51\pm18$&
         $10.6\pm3.8$&$0.64\pm0.23$\\
\cline{2-12}
&\multicolumn{11}{c}{NGC\,6469 --- MS: $1.3\leq m(\ms)\leq3.5$ --- $\rm{Age}=250\pm50$\,Myr}\\
\cline{2-12}
Core&---&---&&---&$9\pm4$&$15\pm3$&&---&---&---&---\\
Halo&---&---&&---&$65\pm31$&$107\pm28$&&---&---&---&---\\
Overall& $1\pm1$&$2\pm2$ &&$1.38\pm0.63$& $74\pm31$&$122\pm29$&&$16\pm13$&$6.3\pm2.8$&
         $13\pm6$&$2.4\pm1.0$\\
\cline{2-12}
&\multicolumn{11}{c}{New\,Cluster\,1 --- MS: $1.1\leq m(\ms)\leq2.1$ --- $\rm{Age}=1.0\pm0.1$\,Gyr}\\
\cline{2-12}
Core&---&---&&---&---&---&&---&---&---&---\\
Halo&---&---&&---&---&---&&---&---&---&---\\
Overall& $7\pm2$&$14\pm4$ &&---& $13\pm8$&$18\pm7$ &&---&---&---&---\\
\hline\hline
\end{tabular}
\begin{list}{Table Notes.}
\item Col.~6: stellar mass stored in the observed MS. Col.~8: mass of the evolved stars added 
to the MS mass extrapolated to 0.08\,\ms.
\end{list}
\end{table*}

\begin{figure}
\resizebox{\hsize}{!}{\includegraphics{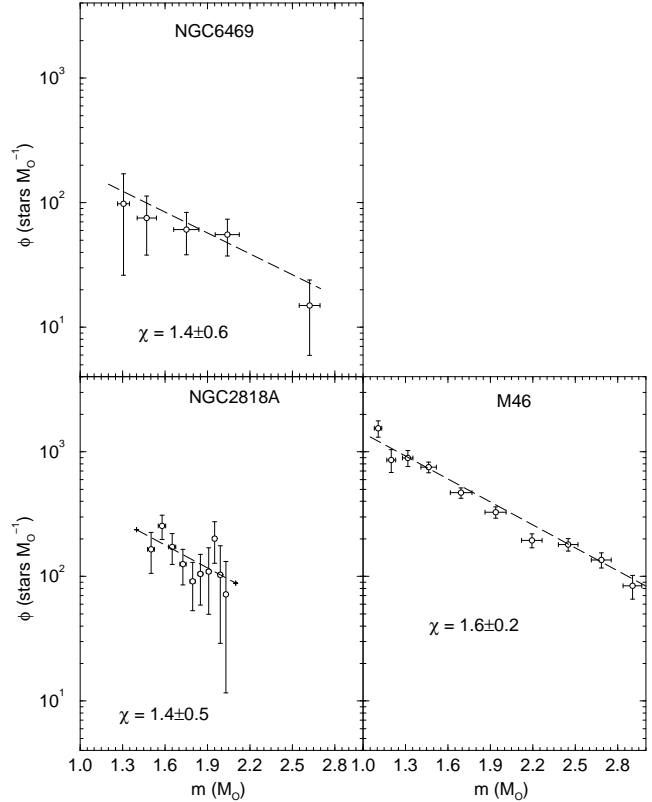}}
\caption[]{Overall mass functions fitted with $\phi(m)\propto m^{-(1+\chi)}$. Within the
uncertainties, the MF slopes are similar to Salpeter's IMF.}
\label{fig7}
\end{figure}

The resulting MFs are shown in Fig.~\ref{fig7}, where fits with the function $\phi(m)\propto m^{-(1+\chi)}$
are included; MF slopes are given in col.~4 of Table~\ref{tab5}. The populous nature of M\,46 allowed
computation of the core and halo MF parameters; for NGC\,2818A and NGC\,6469, only the overall MF was
considered. Within uncertainties, the MFs of the present clusters have slopes similar to that of \citet{Salpeter55}
Initial Mass Function (IMF) ($\chi=1.35$). The core MF of M\,46 appears flatter than the halo's.

Table~\ref{tab5} gives parameters of the target clusters measured in the CMDs and derived from MFs. The
number of evolved stars (col.~2) was obtained by integration of the (background-subtracted) colour-magnitude
filtered luminosity function for stars brighter than the TO. This number multiplied by the stellar mass at
the TO yields an estimate of the mass stored in evolved stars (col.~3). The number and mass of the observed MS
stars (cols.~5 and 6, respectively) were derived by integrati the MFs over the effective MS mass range.

To estimate the total stellar mass we extrapolate the observed MFs down to the H-burning mass limit
($0.08\,\ms$). For masses below the present detection threshold (Table~\ref{tab5}) we base the extrapolation
on \citet{Kroupa2001} universal IMF, in which $\chi=0.3\pm0.5$ for the range $0.08\leq m(\ms)\leq0.5$ and
$\chi=1.3\pm0.3$ for $0.5\leq m(\ms)\leq1.0$. When the present MF slopes are flatter than or similar (within
uncertainties) to Kroupa's, we adopt the measured values of $\chi$. The total (extrapolated MS $+$ evolved)
values of number, mass, surface, and volume densities are given in cols.~7 to 10 of Table~\ref{tab5}.

The mass stored in the observed MS and evolved stars of M\,46 amounts to $\rm M_{obs}\sim1\,500\,\ms$,
compatible with the populous CMD (Fig.~\ref{fig3}). The present observed mass estimate for NGC\,2818A,
$\rm M_{obs}\approx240\,\ms$ agrees with the estimate of \citet{Tad2002}. NGC\,6469, and especially
New\,Cluster\,1, are low-mass OCs, respectively with $\rm M_{obs}\approx120\,\ms$ and $\rm M_{obs}\approx30\,\ms$.
The extrapolated masses are a factor $\sim5$ times larger than the observed ones. As a caveat we note 
that the total mass estimates should be taken as upper limits, since because of dynamical evolution, 
significant fractions of the low-mass content may have been lost to the field.

\section{Comparison with nearby OCs}
\label{CWODS}

At this point it is interesting to compare the structural parameters derived for the present OCs
with those of a reference sample of nearby OCs with ages in the range $70-7\,000$\,Myr and masses
within $400-5\,300$\,\ms\ (\citealt{DetAnalOCs}). To the original reference sample were added the
young OCs NGC\,6611 (\citealt{N6611}) and NGC\,4755 (\citealt{N4755}). Clusters are differentiated
according to total mass (smaller or larger than 1\,000\,\ms). Details on parameter correlation in
the reference sample are given in \citet{DetAnalOCs}.

\begin{figure}
\resizebox{\hsize}{!}{\includegraphics{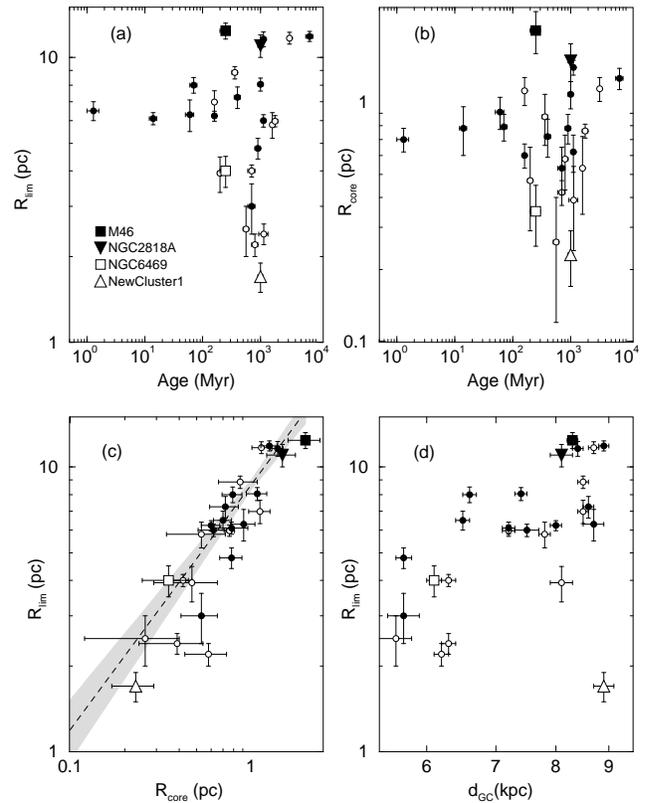}}
\caption[]{Relations involving structural parameters of OCs. Circles: nearby OCs, including
two young ones. Filled symbols: OCs more massive than 1\,000\,\ms.}
\label{fig8}
\end{figure}

In panels (a) and (b) of Fig.~\ref{fig8} we compare limiting and core radii of the present OCs with
those of the reference sample in terms of cluster age. With respect to both kinds of radii, the
bulge-projected NGC\,6469 and the poorly-populated New\,Cluster\,1 appear to be significantly smaller
than the nearby OCs of similar ages, especially in limiting radius. Core and limiting radii in the
reference sample are related by $\rl=(8.9\pm0.3)\times R_{\rm core}^{(1.0\pm0.1)}$ (panel (c)), which
suggests a similar scaling for both kinds of radii, at least for the radii ranges $0.5\la\rc(pc)\la1.5$
and $5\la\rl(pc)\la15$. Within uncertainties, NGC\,2818A, M\,46, NGC\,6469, and New\,Cluster\,1 also
follow that relation. Finally, except for New\,Cluster\,1, the remaining OCs appear to follow the trend
of increasing limiting radii with Galactocentric distance (panel (d)). A similar dependence with \dgc\
holds as well for \rc, because of the relation implied by panel (c).

\section{Discussion}
\label{Discus}

Besides the spatial coincidence, for a physical association between a PN and a star cluster to be
considered as highly probable, at least the radial velocities, reddening values, and distances should
be compatible. In the present work we derive reliable values of reddening, distance from the Sun, core
and limiting radii for the OCs. The angular separation of the PNe NGC\,2818, NGC\,2438, and PK\,6+2.5 with
respect to the OC centres (Table~\ref{tab2}), corresponds to the halo; PK\,167-0.1, on the other hand,
appears to lie near the border of New\,Cluster\,1. In all cases, the mass at the TO of the OC is
consistent with the presence of a PN.

Based on the above criteria, the best candidate to a physical pair is the PN NGC\,2438 with
the OC M\,46. They have comparable radial velocities (Sect.~\ref{Pair1}), and the reddening and 
distance from the Sun for the PN are $\ebv=0.17\pm0.08$ and $\ds\sim1.8$\,kpc (Sect.~\ref{Pair1}),
while for the OC we derive $\ebv=0.10\pm0.03$ and $\ds=1.5\pm0.1$\,kpc. Besides, the presence of
a PN in M\,46 is compatible with the TO mass, $m_{TO}\approx3.5\,\ms$.

Within uncertainties, the present reddening $\ebv=0.10\pm0.03$ for the OC NGC\,2818A is
compatible with the $\ebv=0.18$ and $\ebv=0.28\pm0.15$ (Sect.~\ref{Pair1}) for the PN NGC\,2818. 
Distances from the Sun are in better agreement, $\ds=2.8\pm0.1$\,kpc for NGC\,2818A and
$\ds=2.3 - 2.7$\,kpc for NGC\,2818 (Sect.~\ref{Pair1}). However, the radial velocities do
not agree (Sect.~\ref{Pair1}), which could indicate a projection effect.

PK\,6+2.5 does not appear to be physically associated with NGC\,6469. If this nearby cluster 
(Table~\ref{tab1}) has a circular orbit about the Galactic centre, a small value for its radial 
velocity is expected from its Galactic location. However, the relatively high velocity 
$\rm V_r=68.8\pm1.8\,km\,s^{-1}$ (Sect.~\ref{Pair3}) of PK\,6+2.5 suggests rather a relation to 
the bulge.
 
The distance of 1780\,pc of PK\,167-0.1 (Sect.~\ref{Pair4}) is in excellent agreement with that 
derived for New\,Cluster\,1, $\ds=1.7\pm0.1$\,kpc (Table~\ref{tab1}). This piece of evidence suggests 
physical association, but the cluster radial velocity should be determined for a more robust 
comparison.

\citet{MTL07} comment that most potential Planetary Nebulae in OCs are projected against the cluster
halo. At least in part, this observation can be a direct consequence of the relatively high cluster 
halo/core cross section ratio, $A_h/A_c=(\rl-\rc)^2/\rc^2$. Indeed, from Fig.~\ref{fig8} (panel c) it 
follows that the limiting and core radii are related by $\rl\approx9\,\rc$, which implies that, for a 
typical OC, $A_h/A_c\approx60$. In this sense, random distributions of (spatially unrelated) PNe and OCs 
would result in a significantly larger number of PNe projected against halos than central regions. For 
the 30 spatial coincidences (with separation $\Delta\,R<15\arcmin$), \citet{MTL07} present an estimate 
of a nuclear radius ($r_n$) and measure the angular separation of the PN with respect to the cluster 
centre. Although their $r_n$ is not derived from radial profiles, it probably represents the core radius 
to some degree. For the two additional pairs dealt with in the present paper, the ratio 
of the number of cluster halo/core PNe is $N_h/N_c\approx26/6=4.3$, less than 10\% of that expected of 
random, unrelated distributions. Interestingly, for a King-like cluster characterised by the core and 
limiting radii \rc\ and \rl, the number of member stars in the halo is related to that in the core by
$N_h/N_c=\ln\left[1+(\rl/\rc)^2\right]/\ln(2) - 1$. Thus, for $\rl\approx9\,\rc$, we obtain
$N_h/N_c\approx5.4$, close to the corresponding observed ratio for the PNe of \citet{MTL07}.

{\bf 
We further examine this issue from a different perspective, similar to that employed by \citet{Lund84} 
to study spatial coincidences of WR stars with OCs. We compute the ratio of the PN/OC separation to the 
cluster apparent radius $\rm \chi=\Delta\,R/R_{OC}$, based on the angular diameters given by \citet{Dias02}. 
According to this definition, $\chi$ represents the PN separation as a function of the apparent cluster 
radius. Next we build the surface density distribution of the measured $\chi$ values, i.e, the number of 
spatial coincidences characterised by $\chi$ per cluster area, which is shown in Fig.~\ref{fig9}. To 
investigate the significance of this distribution we first test whether it could result from a random 
spatial distribution of PNe. The 30 PN/OC angular separations in \citet{MTL07}, and the 2 additional
ones  
(Table~\ref{tab2}) not in common with them, are restricted to $\Delta\,R\la55\arcmin$. For each 
of the 32 OCs corresponding to the above spatial coincidences, we randomly take values of $\Delta\,R$ 
uniformly distributed in the range $0\arcmin\la\Delta\,R\la55\arcmin$ and compute $\chi$ for the actual
apparent OC radius. The respective surface density distribution was built for $10^6$ runs, which was 
subsequently normalised to 32 objects. The measured curve presents a conspicuous excess over the simulated 
one for $\chi\la3$, which might indicate a physical relation. Both curves agree for $\chi\ga3$, within the uncertainties. The second test follows \citet{Lund84}: the sign of the Galactic 
latitude ($b$) of each PN is reversed, but keeping the original $\ell$. Based on the new PN coordinates, 
we then searched for an OC that is closer than $\Delta\,R\la55\arcmin$, a condition that was matched for 
20 cases. Within the uncertainties, the corresponding surface density distribution, normalised to 32 
objects (Fig.~\ref{fig9}), coincides with the simulated (random spatial distribution of $\Delta\,R$ values) 
one along the full range of $\chi$. 

The arguments above are based on the apparent cluster radii of \citet{Dias02}. Because they are basically 
visually determined, such radii, however, are significantly underestimated with respect to the limiting 
radii and, especially, the tidal radii. For instance, the OCs in \citet{FaintOCs}, \citet{BB07},
\citet{M52N3960}, and \citet{N4755}, which are similar to those in \citet{MTL07} and were studied with 
the same methods as in the present paper, have a limiting radius $\sim2-3$ times larger than the apparent 
radius given in \citet{Dias02}. Thus, if a similar relation holds for the apparent and limiting radii in the 
present OC sample, the excess in the observed surface density (Fig.~\ref{fig9}) would correspond to PN 
separations $\Delta\,R\la\rl<<\rt$, which again suggests a physical relation.

\begin{figure}
\resizebox{\hsize}{!}{\includegraphics{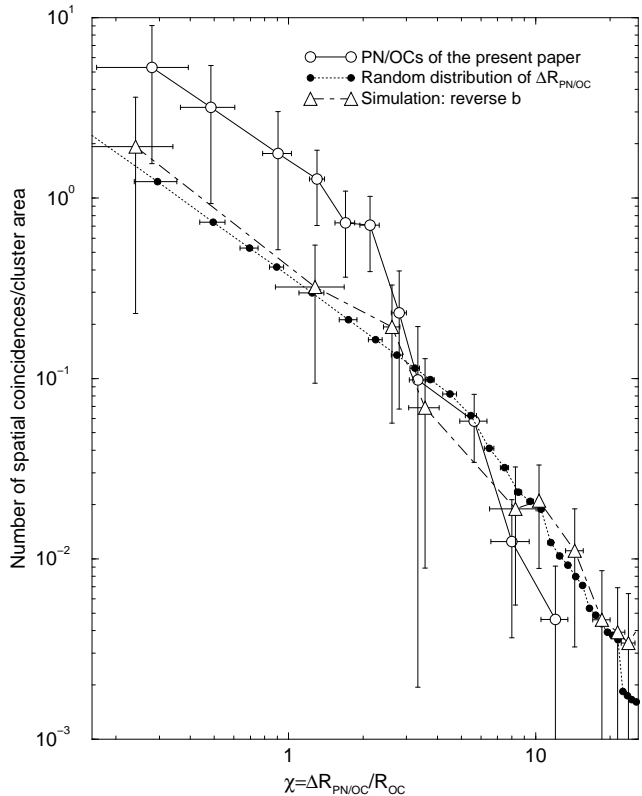}}
\caption[]{Surface density distribution (number of spatial coincidences characterised by
$\chi$ per cluster area). Empty circles: the 32 spatial coincidences in this paper. Filled 
circles: distribution simulated with $10^6$ values of $\Delta\,R$, randomly taken from the 
range $0\arcmin - 55\arcmin$, divided by the actual apparent cluster radii. Empty triangles: 
distribution of the 20 spatial coincidences that result after reversion of the sign of PN's $b$ 
coordinate. The latter two distributons were nomalised to 32 objects. The apparent cluster 
radii (\citealt{Dias02}) are $\sim2-3$ times smaller than the limiting radii.}
\label{fig9}
\end{figure}

}

Despite the many uncertainties associated with the above arguments, {\bf the small number of 
targets and the uncertain PN distances in particular}, it might, nevertheless, be
speculated whether the relatively large fraction of core PNe results from a physical relation.
A cautionary remark is that the remaining spatial coincidences in \citet{MTL07} should be analysed
in detail for a better definition of the core and limiting radii, as well as the fundamental
parameters.
   
\section{Summary and conclusions}
\label{Conclu}

In the present study an open cluster was discovered (New\,Cluster\,1), which is located at 
$\rm\alpha(J2000)=05^h06^m20^s$ and $\delta(J2000)=+39^\circ09\arcmin50\arcsec$. {\bf Projected
on its field, with an angular separation of $4.9\arcmin$ with respect to the custer centre, is 
the PN PK\,167-0.1.} Field-star decontaminated 2MASS photometry was employed in the analysis of 
4 pairs of PN/OC candidates to physical association. We derived accurate ages, distances from the 
Sun, reddening values, and the mass at the turnoff for the clusters. Cluster core and limiting 
radii were determined by means of radial density profiles fitted with a King-like function.

{\bf The values of reddening and distance from the Sun derived in this work suggest that the 
pairs PN/OC NGC\,2818/NGC\,2818A and NGC2438/M46 are compatible with physical associations. 
However, the available radial velocity data for NGC\,2818/NGC\,2818A favour a projection effect.} 
The pair PK\,6+2.5/NGC\,6469 appears to be physically unrelated, since the radial velocity of 
PK\,6+2.5  suggests a bulge membership. The PN PK\,167-0.1 is projected close to the OC 
New\,Cluster\,1, and the similar distances are consistent with physical association. In the 
cases of physical association, the turnoff masses are compatible with the occurrence of PNe.
{\bf It would be important to determine radial velocities for members of the new cluster, to
compare them with that of PK\,167-0.1, in order to further test the physical association
possibility. Besides, to establish the non-association of PK\,6+2.5 with NGC\,6469, radial
veocities of member stars are necessary.}

From the PN/OC angular positions provided by \citet{MTL07}, we estimate that the fraction of 
PNe projected against the central region of clusters, with respect to those in the halo, is
significantly higher than that expected from spatially unrelated distributions of PNe and OCs.
Besides, this fraction agrees with that expected of clusters characterised by a King-like
radial stellar distribution. {\bf In addition, we analysed the distribution of the number of 
spatial coincidences per cluster area built with the 32 cases dealt with in this paper. This
curve was compared to that expected from a spatially-unrelated distribution of PN/OC separations 
and that resulting from inversion of the sign of the $b$ PN coordinates. The observed distribution 
presents a conspicuous excess, for PN/OC separations smaller than the cluster limiting radii, 
over both simulations, which might indicate physical relation. However, we note that these 
results do not take into account the rather uncertain PN distances, and radial velocities. 
Besides, they are based on a relatively small number of spatial coincidences. In this context,
it is essential to analyse in more detail (i.e. with accurate fundamental and structural 
parameters) the whole PN/OC sample of \citet{MTL07} in order to investigate whether the above 
excess of relatively close spatial coincidences can be interpreted as physical association. }

\section*{acknowledgements}
We thank the referee for suggestions.
This publication makes use of data products from the Two Micron All Sky Survey, which
is a joint project of the University of Massachusetts and the Infrared Processing and
Analysis Centre/California Institute of Technology, funded by the National Aeronautics
and Space Administration and the National Science Foundation. This research has made 
use of the WEBDA database, operated at the Institute for Astronomy of the University
of Vienna. We acknowledge support from the Brazilian Institution CNPq.

\label{lastpage}

\begin{thebibliography}{}
 
\bibitem[\protect\citeauthoryear{Bergond, Leon \& Guibert}{2001}]{Bergond2001}
   Bergond, G., Leon, S. \& Guibert, J. 2001, A\&A, 377, 462
   
\bibitem[\protect\citeauthoryear{Bessel \& Brett}{1988}]{BesBret88}
   Bessel, M.S. \& Brett, J.M. 1988, PASP, 100, 1134

\bibitem[\protect\citeauthoryear{Bica \& Bonatto}{2005}]{5LowContr}
   Bica, E. \& Bonatto, C. 2005, A\&A, 443, 465

\bibitem[\protect\citeauthoryear{Bica, Bonatto \& Dutra}{2003}]{CygOB2}
   Bica, E., Bonatto, C. \& Dutra, C.M. 2003, A\&A, 405, 991

\bibitem[\protect\citeauthoryear{Bica, Bonatto \& Dutra}{2004}]{3OpticalCl}
   Bica, E., Bonatto, C. \& Dutra, C.M. 2004, A\&A, 422, 555

\bibitem[\protect\citeauthoryear{Bica, Bonatto \& Blumberg}{2006}]{FaintOCs}
   Bica, E., Bonatto, C. \& Blumberg, R. 2006, A\&A, 460, 83

\bibitem[\protect\citeauthoryear{Bica et al.}{2006}]{GCProp}
   Bica, E., Bonatto, C., Barbuy, B. \& Ortolani, S. 2006, A\&A, 450, 105

\bibitem[\protect\citeauthoryear{Bica, Bonatto \& Camargo}{2007}]{ProbFSR}
   Bica, E., Bonatto, C. \& Camargo, D.. 2007, MNRAS, in press, (astro-ph/0712.0762)
   
\bibitem[\protect\citeauthoryear{Bonatto \& Bica}{2007b}]{BB07}
   Bonatto, C. \& Bica, E. 2007b, MNRAS, 377, 1301

\bibitem[\protect\citeauthoryear{Bonatto, Santos Jr. \& Bica}{2006}]{N6611}
   Bonatto, C., Santos Jr., J.F.C. \& Bica, E. 2006, A\&A, 445, 567

\bibitem[\protect\citeauthoryear{Bonatto, Bica \& Girardi}{2004}]{TheoretIsoc}
   Bonatto, C., Bica, E. \& Girardi, L. 2004, A\&A, 415, 571

\bibitem[\protect\citeauthoryear{Bonatto, Bica \& Santos Jr.}{2005}]{N188}
   Bonatto, C., Bica, E. \&  Santos Jr., J.F.C. 2005, A\&A, 433, 917.

\bibitem[\protect\citeauthoryear{Bonatto \& Bica}{2005}]{DetAnalOCs}
   Bonatto, C. \&  Bica, E. 2005, A\&A, 437, 483

\bibitem[\protect\citeauthoryear{Bonatto \& Bica}{2006}]{M52N3960}
   Bonatto, C. \& Bica, E. 2006, A\&A, 455, 931

\bibitem[\protect\citeauthoryear{Bonatto et al.}{2006a}]{DiskProp}
   Bonatto, C., Kerber, L.O., Bica, E. \& Santiago, B.X. 2006a, A\&A, 446, 121

\bibitem[\protect\citeauthoryear{Bonatto et al.}{2006b}]{N4755}
   Bonatto, C., Bica, E., Ortolani, S. \& Barbuy, B. 2006b, A\&A, 453, 121

\bibitem[\protect\citeauthoryear{Moe \& De Marco}{2006}]{deM06}
   Moe, M. \& De Marco, O. 2006, ApJ, 650, 916


\bibitem[\protect\citeauthoryear{Dias et al.}{2002}]{Dias02}
   Dias, W.S., Alessi, B.S., Moitinho, A. \& L\'epine, J.R.D.
   2002, A\&A, 389, 871

\bibitem[\protect\citeauthoryear{Durand, Acker \& Zijlstra}{1998}]{DAZ98}
   Durand, S., Acker, A. \& Zijlstra, A. 1998, A\&AS, 132, 13

\bibitem[\protect\citeauthoryear{Dutra, Santiago \& Bica}{2002}]{DSB2002}
   Dutra, C.M., Santiago, B.X. \& Bica, E. 2002, A\&A, 383, 219

\bibitem[\protect\citeauthoryear{Eisenhauer et al.}{2003}]{Eisen03}
   Eisenhauer, F., Sch\"odel, R, Genzel, R., Ott, T., Tecza, M., Abuter, R.,
   Eckart, A. \& Alexander, T. 2003, ApJ, 597, L121

\bibitem[\protect\citeauthoryear{Eisenhauer et al.}{2005}]{Eisen05}
   Eisenhauer, F., Genzel, R., Alexander, T. et al. 2005, ApJ, 628, 246

\bibitem[\protect\citeauthoryear{Elson, Fall \& Freeman}{1987}]{EFF87}
   Elson, R.A.W., Fall, S.M. \& Freeman, K.C. 1987, ApJ, 323, 54

\bibitem[\protect\citeauthoryear{Girardi et al.}{2002}]{Girardi2002}
   Girardi, L., Bertelli, G., Bressan, A., et al. 2002, A\&A, 391, 195  

\bibitem[\protect\citeauthoryear{Hurley \& Tout}{1998}]{HT98}
   Hurley, J. \& Tout, A.A. 1998, MNRAS, 300, 977

\bibitem[\protect\citeauthoryear{Kerber et al.}{2002}]{Kerber02}
   Kerber, L.O., Santiago, B.X., Castro, R. \& Valls-Gabaud, D. 2002, A\&A, 390, 121

\bibitem[\protect\citeauthoryear{K\"oppen \& Acker}{2000}]{KA2000}
   K\"oppen, J. \& Acker, A. 2000, ASPC, 211, 151

\bibitem[\protect\citeauthoryear{Kharchenko et al.}{2004}]{KPRSS04}
   Kharchenko, N.V., Piskunov, A.E., R\"oser, S., Schilbach, E. \& Scholz, R.-D.
   2004, AN, 325, 740

\bibitem[\protect\citeauthoryear{Kharchenko et al.}{2005}]{Kharchenko05}
   Kharchenko, N.V., Piskunov, A.E., R\"oser, S., Schilbach, E. \& Scholz, R.-D. 2005,
   A\&A, 438, 1163

\bibitem[\protect\citeauthoryear{King}{1962}]{King1962}
   King, I. 1962, AJ, 67, 471

\bibitem[\protect\citeauthoryear{King}{1966a}]{King66}
   King, I. 1966a, AJ, 71, 64

\bibitem[\protect\citeauthoryear{Kroupa}{2001}]{Kroupa2001}
   Kroupa, P. 2001, MNRAS, 322, 231
   
\bibitem[\protect\citeauthoryear{Lada \& Lada}{2003}]{LL2003}
   Lada, C.J. \& Lada, E.A. 2003, ARA\&A, 41, 57

\bibitem[\protect\citeauthoryear{Lata et al.}{2002}]{Lata02}
   Lata, S., Pandey, A.K., Sagar, R., \&  Mohan, V. 2002, A\&A, 388, 158

\bibitem[\protect\citeauthoryear{Lundstr\"om \& Stenholm}{1984}]{Lund84}
   Lundstr\"om, I. \& Stenholm, B. 1984, A\&AS, 58, 163

\bibitem[\protect\citeauthoryear{Majaess, Turner \& Lane}{2007}]{MTL07}
   Majaess, D.J., Turner, D.G. \& Lane, D.J. 2007, PASP, in press 
   (astro-ph/0710.2900)

\bibitem[\protect\citeauthoryear{Meatheringham, Wood \& Faulkner}{1988}]{MWF88}
   Meatheringham, S.J., Wood, P R. \& Faulkner, D.J. 1988, ApJ, 334, 862

\bibitem[\protect\citeauthoryear{Mermilliod et al.}{2001}]{Merm01}
   Mermilliod, J.-C., Clari\'a, J.J., Andersen, J., Piatti, A.E.
   \& Mayor, M. 2001, A\&A, 375, 30

\bibitem[\protect\citeauthoryear{Mermilliod \& Paunzen}{2003}]{Merm03}
   Mermilliod, J.C. \& Paunzen, E. 2003, A\&A, 410, 511

\bibitem[\protect\citeauthoryear{Meynet, Mermilliod \& Maeder}{1993}]{Meynet93}
   Meynet, G., Mermilliod, J.-C. \& Maeder, A. 1993, A\&AS, 98, 477

\bibitem[\protect\citeauthoryear{Moe \& de Marco}{2006}]{MoeMar06}
   Moe, M. \& de Marco, O. 2006, ApJ, 650, 916

\bibitem[\protect\citeauthoryear{Nilakshi, Pandey, \& Mohan}{2002}]{Nilakshi2002}
   Nilakshi, S.R., Pandey, A.K. \& Mohan, V. 2002, A\&A, 383, 153

\bibitem[\protect\citeauthoryear{Nishiyama et al.}{2006}]{Nishiyama06}
   Nishiyama, S., Nagata, T., Sato, S. et al. 2006, ApJ, 647, 1093

\bibitem[\protect\citeauthoryear{Pauls \& Kohoutek}{1996}]{PK96}
   Pauls, R. \& Kohoutek, L. 1996, AN, 317, 413

\bibitem[\protect\citeauthoryear{Pedreros}{1989}]{Pedreros89}
   Pedreros, M. 1989, AJ, 98, 2146

\bibitem[\protect\citeauthoryear{Phillips}{2004}]{Phillips04}
   Phillips, J.P. 2004, MNRAS, 353, 589

\bibitem[\protect\citeauthoryear{Salpeter}{1955}]{Salpeter55}
   Salpeter, E. 1955, ApJ, 121, 161

\bibitem[\protect\citeauthoryear{Santos Jr., Bonatto \& Bica}{2005}]{M11}
   Santos Jr., J.F.C., Bonatto, C. \& Bica, E. 2005, A\&A, 442,201

\bibitem[\protect\citeauthoryear{Sch\"onberner \& Bl\"ocker}{1996}]{SB96}
   Sch\"onberner, D. \& Bl\"ocker, T. 1996, Ap\&SS, 245,  201
   
\bibitem[\protect\citeauthoryear{Sharma et al.}{2006}]{Sharma06}
   Sharma, S., Pandey, A.K., Ogura, K., Mito, H., Tarusawa, K.,
   \& Sagar, R. 2006, AJ, 132, 1669

\bibitem[\protect\citeauthoryear{Skrutskie et al.}{1997}]{2mass1997}
   Skrutskie, M., Schneider, S.E., Stiening, R., et al. 1997, in {\it The Impact
   of Large Scale Near-IR Sky Surveys}, ed. Garzon et al., Kluwer (Netherlands), 210, 187
   
\bibitem[\protect\citeauthoryear{Soker}{2006}]{Soker2006}
   Soker, N. 2006, ApJ, 645, L57

\bibitem[\protect\citeauthoryear{Tadross et al.}{2002}]{Tad2002}
   Tadross, A.L., Werner, P., Osman, A. \& Marie, M. 2002, NewAst, 7, 553

\bibitem[\protect\citeauthoryear{Tylenda et al.}{1992}]{TASK92}
   Tylenda, R., Acker, A., Stenholm, B. \& K\"oppen, J. 1992, A\&AS, 95, 337

\bibitem[\protect\citeauthoryear{Weidemann}{2000}]{Weidemann00}
   Weidemann, V. 2000, A\&A, 363, 647

\bibitem[\protect\citeauthoryear{Wilson}{1975}]{Wilson75}
   Wilson, C.P. 1975, AJ, 80, 175

\bibitem[\protect\citeauthoryear{Zhang}{1995}]{Zhang95}
   Zhang, C.Y. 1995, ApJS, 98, 659

\bibitem[\protect\citeauthoryear{Zijlstra}{2007}]{Zijlstra07}
   Zijlstra, A.A. 2007, BaltA, 16, 79

\bibitem[\protect\citeauthoryear{\v{Z}i\v{z} Novsk\'y}{1975}]{Ziz75}
   \v{Z}i\v{z} Novsk\'y, J. 1975, BAICz, 26, 248

\end{thebibliography}
\end{document}